\documentclass[aps,amsfonts,amsmath,prd,preprint,nofootinbib]{revtex4}
\pdfoutput=1
\usepackage{graphicx}
\usepackage{bbm}
\usepackage{amsmath}
\usepackage{graphicx}
\usepackage{bm}
\usepackage{amssymb}
\usepackage{latexsym}
\usepackage{graphicx}
\usepackage{dcolumn}
\usepackage{bm}
\usepackage{mathrsfs}

\def\be{\begin{equation}}
\def\ee{\end{equation}}
\def\ba{\begin{eqnarray}}
\def\ea{\end{eqnarray}}
 
\begin{document}

\title{Topological Defects from the Multiverse}
\author{Jun Zhang$^a$, Jose J. Blanco-Pillado$^{b,c}$, Jaume Garriga$^d$ and Alexander Vilenkin$^a$}

\address{$^a$Institute of Cosmology, Department of Physics and Astronomy,\\ 
Tufts University, Medford, MA 02155, USA,\\
$^b$Department of Theoretical Physics,\\ 
University of the Basque Country UPV/EHU, 48080 Bilbao, Spain\\
$^c$ IKERBASQUE, Basque Foundation for Science,  48013, Bilbao, Spain\\
$^d$Departament de Fisica Fonamental i Institut de Ciencies del Cosmos,\\
Universitat de Barcelona, Marti i Franques, 1, 08028, Barcelona, Spain}

\begin{abstract}
Many theories of the early universe predict the existence of a multiverse where bubbles continuously nucleate giving rise to observers in their interior. In this paper, we point out that topological defects of several dimensionalities will also be produced in de Sitter like regions of the multiverse. In particular, defects could be spontaneously nucleated in our parent vacuum.  We study the evolution of these defects as they collide with and propagate inside of our bubble. We estimate the present distribution of defects in the observable part of the universe. The expected number of such nearby defects turns out to be quite small, even for the highest nucleation rate.  We also study collisions of strings and domain walls with our bubble in our past light cone. We obtain simulated full-sky maps of the loci of such collisions, and find their angular size distribution. Similarly to what happens in the case of bubble collisions, the prospect of detecting any collisions of our bubble with ambient defects is greatly enhanced in the case where the cosmological constant of our parent vacuum is much higher than the vacuum energy density during inflation in our bubble.
\end{abstract}
\maketitle

\section{Introduction}

A wide variety of particle physics models predict the existence of a multiverse, where bubbles of different vacua nucleate and expand in an eternally inflating background.  Occasionally bubbles collide, and observers inside 
the bubbles can search for the imprints left by these collisions on the cosmic background radiation (CMB).  Detecting such imprints would be tremendously important, as they would provide a direct evidence for the existence of a multiverse. The dynamics of these collisions have been studied analytically as well as numerically over the years~\cite{Hawking82, Wu83,JJ04} and the analysis of their observational signatures has now become an active area of research; see, e.g.,~\cite{GGV,Aguirre1,Kleban1,Freivogel,Aguirre2,Salem10,Kleban2,Johnson} and references therein.

In the present paper, we shall discuss another potentially observable manifestation of a multiverse.  It has been shown in \cite{Basu} that topological defects, like spherical domain walls and circular loops of cosmic string can be spontaneously  produced in a de Sitter-like universe. The initial radii of walls and strings are close to the de Sitter horizon. Once they are formed, they are stretched by the inflationary expansion, resulting in a scale-invariant size distribution of walls and/or strings.  If indeed we live in an expanding bubble, our bubble will collide with topological defects nucleating in the inflating parent vacuum.  Any such collisions that occurred within our past light cone can give rise to potentially observable effects.  

The physical properties of the defects depend on the embedding vacuum and may change significantly as the defects cross from the parent vacuum to the bubble interior.  In fact, matching defect solutions may not even exist inside the bubble.  Collisions of defects with our bubble would then result in the disintegration of the defects.  The opposite limit is when the fields making up the defects do not interact with the field of the bubble, except gravitationally.  Then the defects crossing from parent vacuum into the bubble will only respond to the change of spacetime geometry. 
Walls and strings encountered by our bubble will then remain, at least partially, in the bubble interior and may be observable directly.  These defects, originating in the false vacuum, can in principle be much heavier than strings and walls that can be formed in phase transitions inside our bubble.

Our goal in this paper is to study the dynamics of the intruding defects and to estimate the number of potentially observable collisions. We shall focus on cosmic strings, but most of our discussion will apply to domain walls as well. In the next Section we specify our assumptions and approximations and find the conditions for a string loop or a spherical domain wall to collide with the bubble. In Section III, we study the propagation of the defects that do get inside the bubble and find their asymptotic shape. The spatial distribution of defects is discussed in Section IV, and the number of potentially observable collision events is estimated in Section V.  The distribution of collision regions on the observer's sky is discussed in Section VI, although the investigation of the nature of observable effects due to collisions is left for future study. Our conclusions are summarized in Section VII.  Some technical details of the calculations are given in the Appendices.

\section{Bubble-defect collisions}

\subsection{Bubble spacetime}

In this section, we shall discuss the conditions for a string loop (or domain wall) nucleated outside of our bubble to collide with the bubble. To set up the problem, we assume the spacetime outside the bubble, which we shall refer to as the false vacuum, to be a de Sitter space with a constant expansion rate $H_F$. It can be partially covered by flat chart metric
\ba
ds^2 = dt^2 - H_F^{-2} e^{2H_Ft} \left(dr^2+r^2 d\Omega_2^2 \right),
\label{dS}
\ea 
where $d\Omega_2^2=d\theta^2+\sin^2\theta d\phi^2$.  It will also be convenient to use Cartesian coordinates $(x,y,z)$, related to $(r,\theta,\phi)$ by
\ba
dr^2+r^2 d\Omega_2^2 = dx^2+dy^2+dz^2.
\ea

The spacetime inside the bubble has the geometry of an open FLRW universe,
\ba
ds^2 = d\tau^2 - a^2\left(\tau\right) \left(d\xi^2+\sinh^2\xi d\Omega_2^2\right).
\ea
We shall refer to it as true vacuum and approximate it by a de Sitter space with an expansion rate $H_T \leq H_F$,
\ba
a\left(\tau\right)= H_T^{-1} \sinh (H_T \tau) .
\ea
The parameter $H_T$ has the meaning of the slow-roll inflation rate inside the bubble at the early stages of its evolution.  At later times, the form of $a(\tau)$ should be different, corresponding to radiation and matter eras.  However, we shall see that the shape of strings gets frozen in comoving coordinates soon after they enter the bubble, so the late-time behavior of $a(\tau)$ is unimportant for our consideration.  

We shall disregard the gravitational effects of both the bubble wall and the defect.  Furthermore, we shall assume that the bubble wall thickness and its initial radius at nucleation are both small compared to the de Sitter horizon $H_F^{-1}$.  Then the worldsheet of the wall is well approximated by the forward light cone,
\ba
r=1-e^{-H_F t},
\label{wallworldsheet}
\ea
where we have chosen the nucleation center to be at the the origin of the flat slicing coordinates given by Eq. (\ref{dS}), namely, $r=t=0$.

\subsection{Defect nucleation}

The nucleation rate of strings and walls is 
\be
\Gamma \propto e^{-S_E} ,
\label{Gamma}
\ee
where $S_E$ is the Euclidean action of the corresponding instanton.  In the case of strings, the Euclidean worldsheet in the thin wall limit has the form of a maximal 2-sphere of radius $H_F^{-1}$ embedded in a 4-sphere of the same radius, and the instanton action is 
\be
S_E = 4\pi\mu H_F^{-2},
\label{SE}
\ee
where $\mu$ is the string tension.  The string worldsheet after nucleation is obtained as an analytic continuation of this instanton\footnote{One can show that these are solutions of the Nambu-Goto equations of motion of strings in a de Sitter background.}. For a loop centered at $x=r_n$ on the $x$-axis, it is given by \cite{Basu} 
\ba
(x - r_n)^2 + y^2 +z^2 =  e^{-2H_Ft_n} + e^{-2H_Ft},~~z=\left(x-r_n\right)\tan\alpha,
\label{stringworldsheet}
\ea
where $t_n$ is a constant parameter related to the nucleation time and $\alpha$ is the angle between the plane of the loop and the $xy$-plane.  The point with coordinates $(t,x,y,z)=(t_n,r_n,0,0)$ is called the nucleation center of the loop.  It has the important property that its future light cone asymptotically approaches the worldsheet of the loop at $t\to \infty$.

The physical radius of the loop at time $t$ is
\ba
R(t) = H_F^{-1}\left(e^{2H_F(t-t_n)} +1\right)^{1/2} .
\label{Rt}
\ea
It asymptotes to $H_F^{-1}$ at $t\to -\infty$ and expands as $R(t)\propto \exp(H_F t)$ at $t\gg t_n$.  
The part of the worldsheet at $t\to -\infty$ is unphysical and should be cut off at some $t\sim t_n$.  
However, unlike the case of bubble nucleation, the loop radius in (\ref{Rt}) changes monotonically with time, so there is no natural choice for the moment of nucleation.  The proposal of Ref.~\cite{Basu} is that the loop is formed at the time $t_f$ such that the classical action, evaluated from $t\to -\infty$ to $t_f$, is $S\sim 1$.  This gives 
\be
t_n - t_f \sim H_F^{-1}\ln S_E, 
\label{tf}
\ee
where $S_E$ is the Euclidean action (\ref{SE}). The semiclassical description of string nucleation requires that $S_E\gtrsim 1$.  On the other hand, we shall be interested in the situation when $S_E$ is not too large, so that the nucleation rate is not too strongly suppressed.  Then Eq.~(\ref{tf}) suggests that $t_f$ is a few Hubble times earlier than $t_n$.

Another issue related to the worldsheet cutoff at some time $t_f$ is that such a cutoff selects a reference frame where the loop nucleates.  Analysis in Ref.~\cite{SugumidS} suggests that this frame is determined, at least in part, by the initial conditions at the onset of inflation.\footnote{The frame of loop nucleation may also be partly determined by the rest frame of the detector used to observe the loop \cite{Sugumi1,Sugumi2}.  In our case, however, the role of the detector is played by the bubble, which does not have a rest frame.  The loop collides with different parts of the bubble, moving at different speeds.  The results of Refs.~\cite{SugumidS,Sugumi1,Sugumi2} cannot therefore be directly applied to this case.  This issue requires further study.  In the meantime, here we adopt the simple model outlined in the text.}  The existence of a preferred frame breaks the de Sitter invariance of the false vacuum.  This memory of the initial state extends indefinitely into the future; it is a manifestation of the so-called persistence of memory effect \cite{GGV}. We shall assume that the coordinate system in (\ref{dS}) is chosen so that the loops nucleate on constant-$t$ surfaces.

The number of string loops of radius $R$, whose centers are located in a 3-volume element $dV$ on any constant-time surface, is given by \cite{Basu} 
\ba\label{dN}
dN = \lambda  \frac{dR}{R^4} dV,
\ea
where we have defined $\lambda\equiv \Gamma H_F^{-4}$ and we assumed a lower cutoff at $R\sim H_F^{-1}$. The shape of the distribution near the cutoff depends on the details of the nucleation process, so strictly speaking, Eq.~(\ref{dN}) applies only for $R\gg H_F^{-1}$.

The description of domain wall nucleation is very similar.  The Euclidean worldsheet in this case is a 3-sphere of radius $H_F^{-1}$ and the instanton action is
\be\label{SE2}
S_E=2\pi^2 \sigma H_F^{-3},
\ee
where $\sigma$ is the wall tension.  The wall radius as a function of time and the time of nucleation $t_f$ are given by the same Eqs.~(\ref{Rt}), (\ref{tf}).

True vacuum bubbles may interfere with defect nucleation.  As we mentioned in the Introduction, defect solutions may not even exist inside the bubbles.  Here we shall assume that defects cannot be formed if they overlap with any bubbles at the time of their formation.\footnote{Even if string solutions do not exist in true vacuum, open string segments could still nucleate outside the bubble with their ends attached to the bubble wall.  Similarly, strings that can exist inside but not outside the bubble can nucleate with their ends attached to the bubble wall on the interior side.  Nucleation and evolution of such defects will be discussed elsewhere.}  The defect radius at the time of formation is $R_f\approx H_F^{-1}$, and the nucleation process can be pictured as quantum tunneling from $R=0$ to $R=R_f$ \cite{Basu}.  We shall assume that defects can be formed only if they do not encounter any bubbles in the course of the tunneling -- that is, if there are no bubbles in the defect interior at $t=t_f$.  On the other hand, once a defect is formed, true vacuum bubbles can nucleate in its interior.

\subsection{Conditions for bubble-defect collision}
\label{colorcodesec}

Nucleating loops of string will have random orientations, with the angle $\alpha$ in (\ref{stringworldsheet}) ranging from $0$ to $\pi/2$.  For simplicity, here we only consider the case of $\alpha=0$.
In the case of domain walls, the wall radius satisfies the same Eq.~(\ref{Rt}) as the radius of a string loop.  The great circle on the wall, which lies in the plane passing through the center of the bubble, evolves in exactly the same way as a string loop with an inclination angle $\alpha=0$.  For definiteness, in this and the following section we shall focus on strings, but the results will also apply to domain walls and will actually be more complete for walls (since we only consider the case of $\alpha=0$).

In spherical coordinates $(t,r,\theta,\phi)$, the worldsheet of a string loop with $\alpha=0$ can be written as
\ba\label{loop1}
r^2 - \frac{2\sin\chi_n\cos\phi}{\cos \eta_n+\cos\chi_n}r+\frac{\cos\eta_n-\cos \chi_n}{\cos \eta_n+\cos\chi_n}= e^{-2H_Ft},~\theta=\pi/2 ,
\ea
where, for later convenience, we have re-expressed the coordinates $t_n$ and $r_n$ of the nucleation center in terms of its conformal coordinates in the closed slicing of de Sitter space defined by
\ba
&&e^{-H_F t_n} =-\frac{\sin \eta_n}{\cos \eta_n + \cos \chi_n} , \nonumber\\
&&r_n=\frac{\sin \chi_n}{\cos \eta_n + \cos \chi_n} ,
\ea
with $0 \le \chi_n \le \pi$ and $-\pi \le \eta_n \le 0$.  Assuming that the nucleation center is in the part of de Sitter space covered by the flat-slicing coordinates (\ref{dS}), we must have $\chi_n \le \eta_n + \pi$.  All coordinate systems that we use in this paper and the relations between them are summarized in Appendix A.

To derive the conditions for a string loop to cross the bubble wall (\ref{wallworldsheet}), we shall focus on the $x$-axis, $y=z=0$, where we can find the nearest point and the most distant point on the loop with respect to the bubble center.  The intersections of this axis with the string loop and with the bubble are
\ba
&&x_{S\pm}\left(t\right) = \frac{\sin \chi_n}{\cos \eta_n + \cos \chi_n} \pm \sqrt{e^{-2H_Ft}+\frac{\sin^2\eta_n}{\left(\cos \eta_n + \cos \chi_n\right)^2}},\label{xs} \\
&&x_{B\pm}\left(t\right)=\pm \left(1-e^{-H_F t}\right),\label{xb}
\ea
where the subscript $S$ denotes the string loop, $B$ denotes the bubble, and $\pm$ denote the right and left parts of  the string loop/bubble respectively. For definiteness, we shall assume that the loop nucleation center lies on the positive side of the $x$-axis, that is, at $x=r_n >0$. Then, in order to have a collision, we need 
\ba
\mid x_{S-}\left(t \to \infty \right) \mid ~ <~ 1,
\ea
or $-\pi/2 < \chi_n+\eta_n<\pi/2$, and the string loop will completely enter the bubble if it also satisfies\footnote{Since the loop nucleation center is assumed to be at $x>0$, it is obvious that we always have $x_{S+} >0$.}
\ba
x_{S+}\left(t \to \infty \right) < 1,
\ea
or $\chi_n-\eta_n < \pi/2$. 
The collision starts on the right side of the bubble (that is, on the positive half of the x-axis) if
\ba\label{conleftright}
x_{S-}\left(0\right) \ge 0,
\ea
or $\chi_n \ge \pi/2$; otherwise, the collision, if there is one, will start on the left side of the bubble. 

Finally, by solving $x_{S-} = x_{B\pm}$, we find the time of the collision,
\ba\label{tc}
t_{c\pm}= \frac{1}{H_F}\log\left[1+\frac{\cos \chi_n}{\cos\eta_n \mp \sin \chi_n}\right].
\ea
There is always a solution of $x_{S-} = x_{B\pm}$. However, for $t_{c\pm}$ to have physical meaning, we need $t_{c+} > 0$ if the collision is from the left or $t_{c-}>0$ if the collision is from the right, since the bubble worldsheet described by Eq. (\ref{xb}) with $t<0$ is unphysical.

\begin{figure}
 \includegraphics[width=0.6\textwidth]{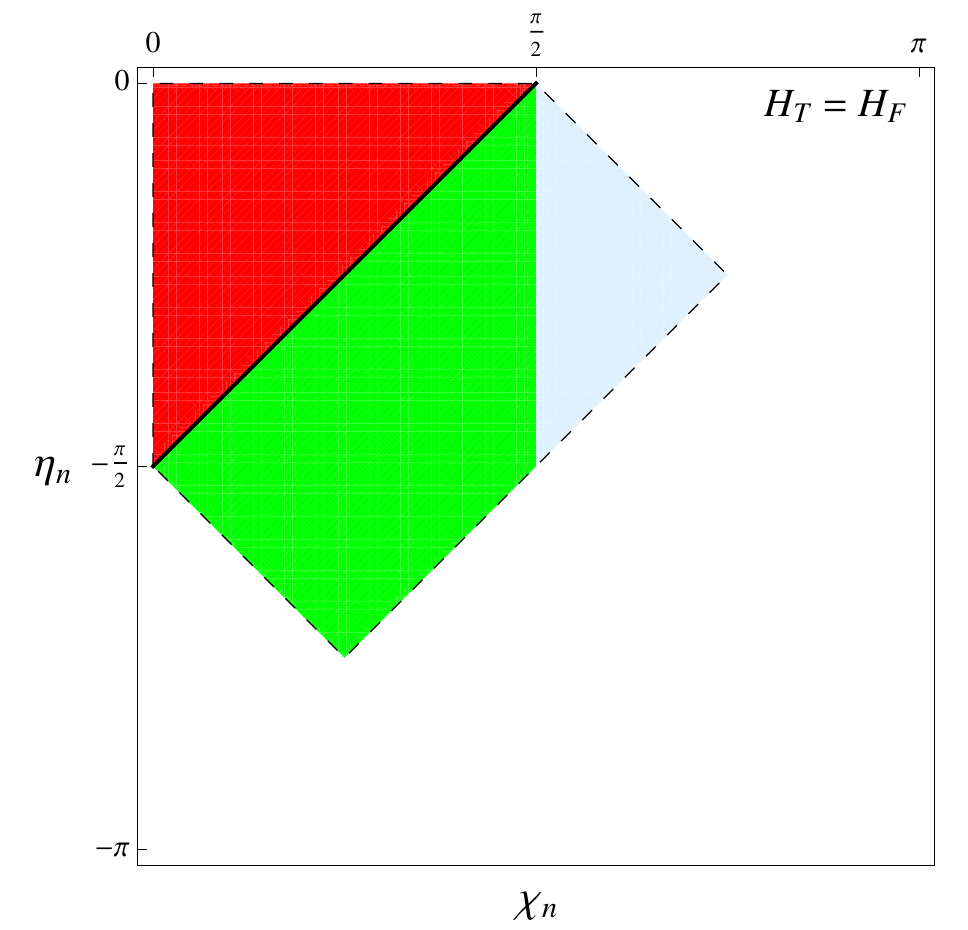}
\caption{The different regions of the parameter space of $(t_n, r_n)$ shown over the conformal diagram of de Sitter space. String loops with nucleation centers in the blue and green regions will partially enter the bubble, and string loops with nucleation center in the red region will completely enter the bubble. String loops with nucleation centers in the green and red regions will hit the bubble from the left, and string loops with nucleation center in the blue region will hit it from the right.}
\label{fig:para}
\end{figure}

To summarize, we project all the above constraints into the parameter space of $(\eta_n, \chi_n)$ in Fig. \ref{fig:para}.  The bubble nucleation center is at $(\eta_n, \chi_n) = (-\pi/2,0)$, and the bubble wall is approximated by the future light cone of this point.  String loops whose nucleation centers are inside the bubble (that is, in {the red area} in the diagram) will completely enter the bubble.  This follows from the fact that the future light cones of such nucleation centers are completely contained within the bubble.  On the other hand, the future light cones of nucleation centers in {the green and blue areas} are only partially contained in the bubble.  The corresponding loops will therefore only partially enter the bubble.
String loops with nucleation centers in the blue area of the diagram hit the bubble from the right, while loops with nucleation centers in the red and green areas hit it from the left.  In the latter case, the loop must encircle the bubble; hence, it has to be formed at $t<0$, prior to bubble nucleation.

We emphasize that loops can form in the false vacuum even when their nucleation centers are in the true vacuum inside the bubble.  For example, the loop nucleation center can be in the red area of the diagram, while the loop formation time, which is earlier than $t_n$, can be at $t_f < 0$ (that is, $\eta_f < -\pi/2$).  The loop is then formed prior to the bubble nucleation.  When the bubble forms, it is initially encircled by the loop, and as it expands, the loop is eventually completely engulfed by the bubble.  

We note that the division of spacetime into different regions in Fig.~\ref{fig:para} is invariant under a class of de Sitter boosts. This is obvious for the boundary between {the red and green regions}, which corresponds to the bubble wall, but may need a demonstration for the boundary separating {the blue and green regions}. According to Eq. (\ref{conleftright}), this boundary is given by $\chi_n=\pi/2$, or
\ba
r_n = \sqrt{1+e^{-2H_Ft_n}}.
\ea 
Using the embedding coordinates $(X,Y,Z,W,V)$ defined in Appendix A, this can be rewritten as 
\ba
W=0.
\label{W}
\ea
On the other hand, the bubble wall lies at $W = H_F^{-1}$.  de Sitter transformations that leave the wall invariant correspond to $5D$ Lorentz boosts in the $X$, $Y$ and $Z$ directions,  
\ba\label{boost}
&&V'=\frac{V-\beta X_i}{\sqrt{1-\beta^2}} \nonumber \\
&&X'=\frac{X_i-\beta V}{\sqrt{1-\beta^2}} \nonumber \\
&&W'=W,
\ea
where $\beta$ is the boost constant.  These transformations do not affect the $W$ coordinate and thus leave the boundary (\ref{W}) invariant. (Boosts in the $W$ direction change the nucleation time parameter $t_n$ of the loop.)

The boost invariance of the boundary between blue and green regions can also be understood geometrically.  Defects with nucleation centers in blue and green regions hit the bubble respectively from the right and from the left.  By continuity, defects with nucleation centers on the boundary between these regions must hit the bubble at the moment of its formation.  Worldsheets of such defects should pass through the origin of the light cone representing the bubble.  This is a geometric fact which is invariant under boosts.

\section{String propagation inside the bubble}

We would now like to study the dynamics of the string after it enters the bubble.  We assume that the fields making up the string do not interact with the field of the bubble, so the string responds only to the change of spacetime geometry.  We shall start with the case of $H_T=H_F$, and then extend the analysis to the case of $H_T \le H_F$.

\subsection{$H_T=H_F$}

In this case, the whole spacetime is a pure de Sitter space; hence the worldsheet of the string loop can be described by Eq.~(\ref{loop1}) both inside and outside the bubble. The only difference is that the observer inside the bubble uses the open-slicing coordinates, which relate to the flat coordinates used earlier by
\ba
&&\cosh H\tau=\cosh Ht - \frac12 e^{Ht}r^2 \nonumber \\
&&\sinh H\tau \sinh \xi = e^{Ht} r,
\ea
where $H_F=H_T=H$. Thus, according to the observer inside the bubble, the worldsheet of the string found earlier in Eq. (\ref{loop1}) takes the form
\ba\label{loop2}
\cos \eta_n \cosh \xi \tanh H \tau - \sin \chi_n \cos \phi \sinh \xi \tanh H\tau -\cos \chi_n =0 .
\ea

\begin{figure}
 \includegraphics[width=1\textwidth]{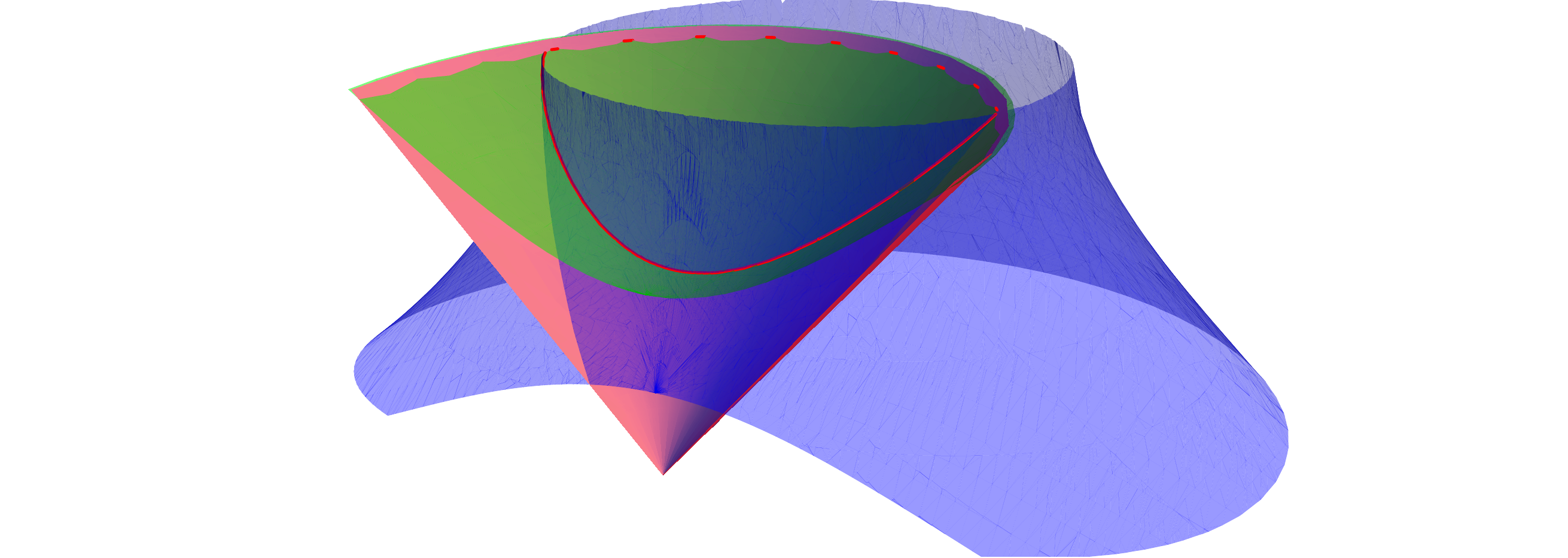}
\caption{Intersecting worldsheets of the bubble wall and of a nucleated string loop shown in the conformal coordinates. The pink surface is the worldsheet of the bubble, the green surface is a constant time surface in the open-slicing coordinate system inside the bubble, and the blue surface is the worldsheet of the string loop. The size of the string loop in conformal coordinates shrinks with time, but the actual physical size grows with time. Highlighted by the red line is the intersection between the green and blue surfaces,  showing the shape of the string at a certain time. The string has infinite length and a fixed asymptotic angle at spatial infinity.}
\label{fig:inter}
\end{figure}

As illustrated in Fig.~\ref{fig:inter}, the shape of the string can be found by setting $\tau ={\rm const}$ in this equation. At $\tau\gg H^{-1}$, $\tanh H\tau \approx 1$, and thus
the string freezes in the comoving coordinates within a few Hubble times after it enters the bubble, approaching the asymptotic shape
\ba\label{xi2}
\cos \eta_n \cosh \xi - \sin \chi_n \cos \phi \sinh \xi -\cos \chi_n =0.
\ea
Eq.(\ref{xi2}) has two solutions for $\xi(\phi)$,
\ba \label{xipm}
e^{\xi_{\pm}} = \frac{2\cos \chi_n \pm \sqrt{1-2\cos2\eta_n + \cos2\phi + 2 \cos 2\chi_n \sin^2\phi}}{2\cos \eta_n -2\cos\phi \sin \chi_n},
\ea
which are related by $\xi_{+}\left(\phi\right) = -\xi_{-}\left(\phi + \pi\right)$; the physical solution is chosen by requiring $\xi \ge 0$. Thus, strings nucleating in the blue region are described by $\xi_-$, and strings nucleating in the green region are described by $\xi_+$. For some of the strings nucleating in the red region, we need the positive parts of both $\xi_+$ and $\xi_-$. 

\begin{figure}
 \includegraphics[width=0.6\textwidth]{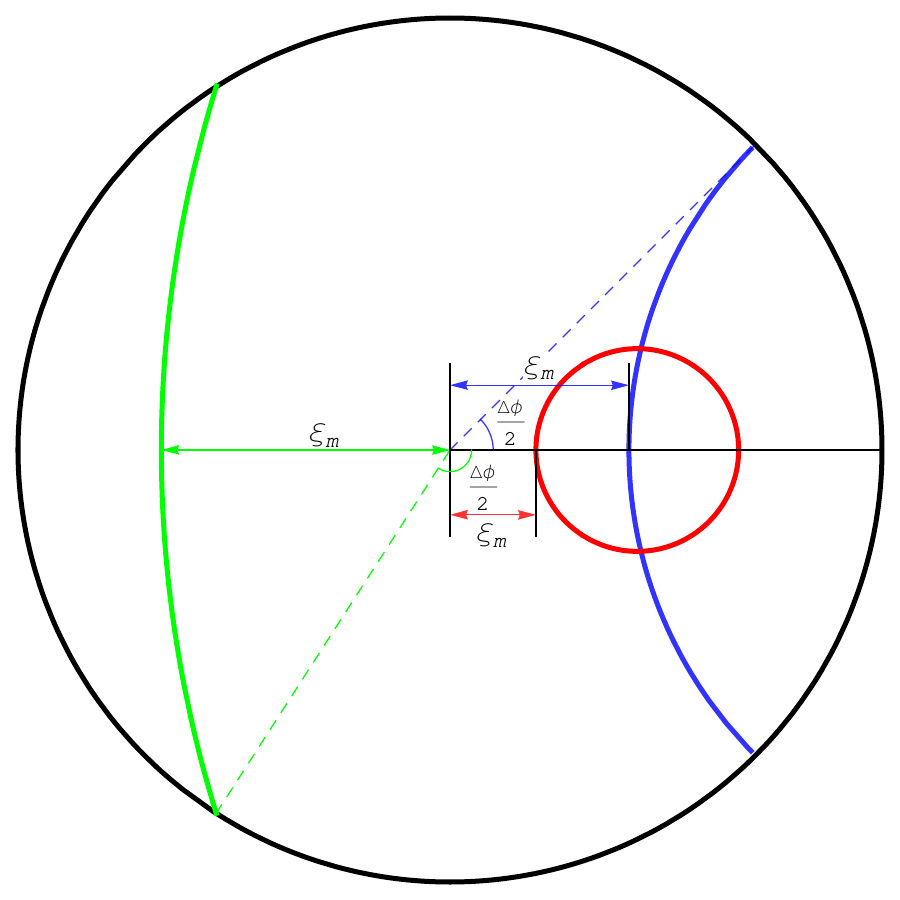}
\caption{The shape of strings at $\tau \to \infty$ projected on the Poincar\'e disk. We plotted the shape of 3 different kinds of loops and labeled the corresponding $\xi_m$ and $\Delta\phi$. 
The parameters $(\eta_n, \chi_n)$ are $(-\pi/4,\pi/2)$ for the blue loop, $(-\pi/8,\pi/4)$ for the red loop and $(-5\pi/8,\pi/4)$ for the green loop.}
\label{fig:pd}
\end{figure}

We can define the minimal distance from the string to the origin as $\xi_-$ at $\phi = 0$:
\ba\label{xim}
\xi_{m}= \log\left[ \frac{\cos \chi_n+\sin \eta_n}{\cos \eta_n -\sin \chi_n}\right]= \log\left[\frac{e^{Ht_n}\left( 1+r_n\right)-1}{e^{Ht_n}\left(1- r_n\right)+1}\right],
\ea
where we also give the expression of $\xi_m$ in terms of $t_n$ and $r_n$ for later use. According to the relation between $\xi_+$ and $\xi_-$, a negative $\xi_m$ indicates that the minimal distance appears on the left half of the x-axis. To clarify the meaning of $\xi_m$, it is helpful to project the strings at $\tau \to \infty$ onto the Poincar\'e disk. Introducing $z_P= \tanh \frac{\xi}{2}$ and $\phi_P=\phi$, we can project the plane of $\theta = \pi/2$ at constant time $\tau$ onto a unit disk. 
Then a string at $\tau \to \infty$ has the form
\ba
z_P = z_0 \cos \phi_P + \sqrt{R_P^2-z_0^2 \sin^2 \phi_P},
\ea
which is a circle on the Poincar\'e disk. Here, we have defined $z_0=\frac{\sin \chi_n} {\cos \chi_n + \cos \eta_n}= r_n$ and $R_P=-\frac{\sin\eta_n}{\cos \chi_n + \cos \eta_n}=e^{-H_Tt_n}$. In Fig. \ref{fig:pd}, we show some examples of different kinds of loops and mark the corresponding distances $\xi_m$. The asymptotic shape of the loop depends on the location of its nucleation center $(\eta_n, \chi_n)$.

\begin{figure}
  \includegraphics[width=0.8\linewidth]{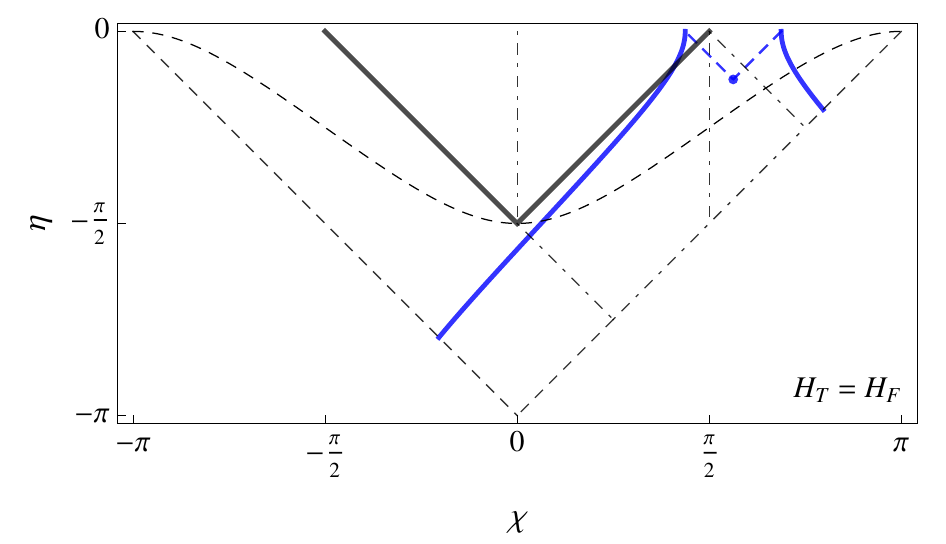}%
\caption{An example of string worldsheet with nucleation center in the blue region of the conformal diagram (Fig.~\ref{fig:para}) in the case when $H_T = H_F$.  The thick blue and black solid lines are the worlsheets of the string loop and of the bubble wall, respectively.  The loop is formed outside of the bubble and then partially enters the bubble.  The blue dashed lines denote the light cone of the loop nucleation center, which is represented by a blue dot. The black dashed lines indicate the surfaces of $t = 0$ and $t = -\infty$ of the flat slicing. Finally, the dot-dashed lines show the boundaries between different parameter space regions of Fig.~\ref{fig:para}.}
\label{fig:cfdb}
\end{figure}

For a loop nucleated in {green or blue} regions in Fig. \ref{fig:para}, the string will appear to have infinite length, with its two asymptotes becoming straight and spanning a fixed angle from the origin. This angle can be found by sending $\xi \to \infty$,
\ba\label{dphi}
\Delta\phi=2\arccos\left[\frac{\cos\eta_n}{\sin\chi_n}\right]= 2\arccos\left[\frac{ \left(1+ r_n^2\right)-e^{-2 H t_n}}{2 r_n}\right].
\ea

\begin{figure}
  \includegraphics[width=0.8\linewidth]{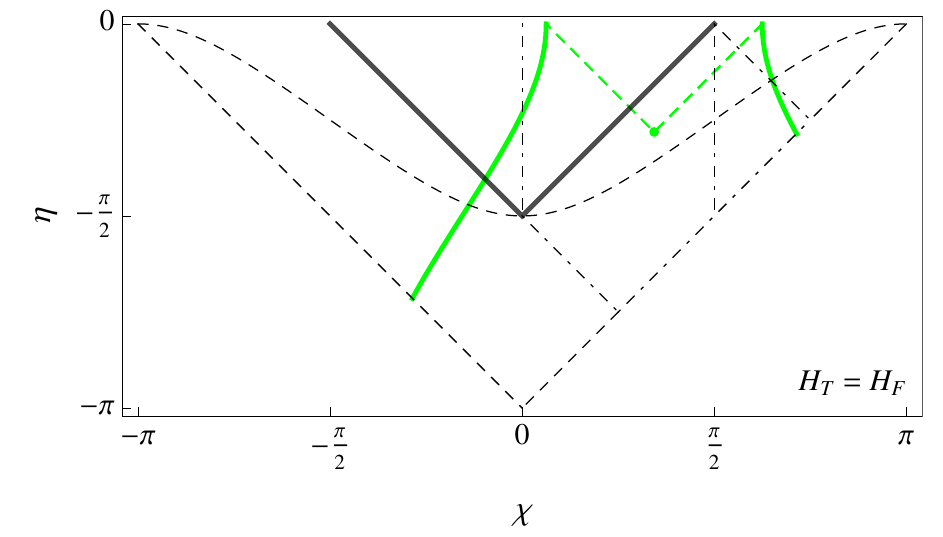}%
 \caption{An example of string worldsheet with nucleation center in the green region of the conformal diagram 
 in the case when $H_T = H_F$.  The loop initially encircles the bubble and then partially enters the bubble. }
 \label{fig:cfdg}
\end{figure}

\begin{figure}
  \includegraphics[width=0.8\linewidth]{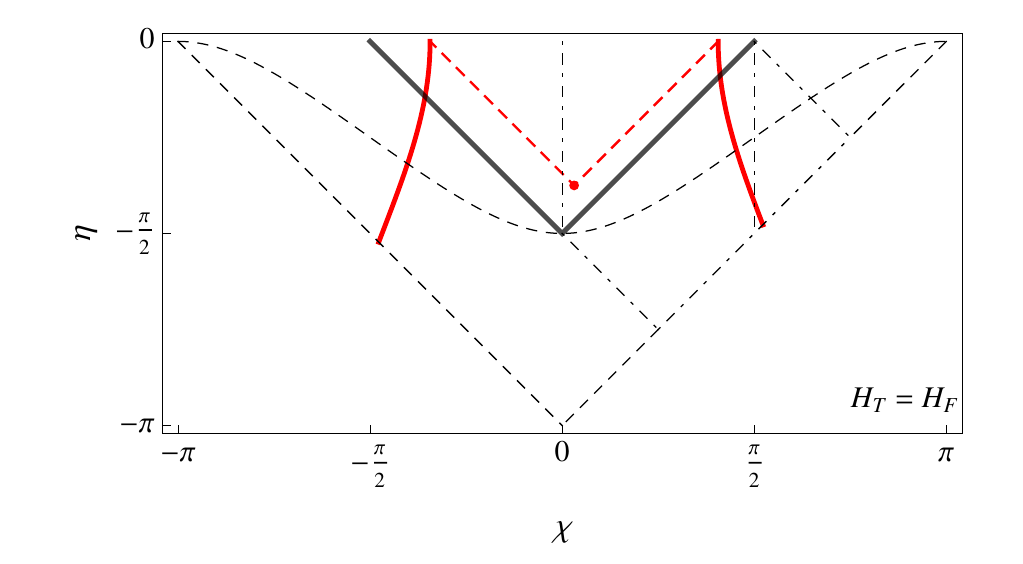}%
\caption{An example of string worldsheet with nucleation center in the red region of the conformal diagram in the case when $H_T = H_F$. The loop initially encircles the bubble and ends up completely inside the bubble.}
\label{fig:cfdr}
\end{figure}

For a nucleation center {in the red region} of Fig. \ref{fig:para}, the loop initially surrounds the origin, then shrinks and finally freezes in comoving coordinates, but physically always expands. At $\tau \to \infty$, we can find the maximal distance from the origin to the loop:
\ba\label{ximaxloop}
\xi_{a}= -\log\left[ \frac{\cos \chi_n+\sin \eta_n}{\cos \eta_n + \sin \chi_n}\right]= -\log\left[\frac{e^{Ht_n}\left(1- r_n\right)-1}{e^{Ht_n}\left( 1+r_n\right)+1}\right].
\ea
Some examples of loop worldsheets with different nucleation centers are shown in Figs.~\ref{fig:cfdb}-\ref{fig:cfdr}\footnote{We use a $1+1$ dimensional version of the conformal diagram for de Sitter in these figures in order to be able to illustrate both sides of the string worldsheet.}.

\subsection{$H_T < H_F$}

In this case, we cannot use Eq.~(\ref{loop1}) after the string enters the bubble.  In order to extend the string worldsheet into the bubble, we use our assumptions that the bubble wall gravity can be neglected and that the initial size of the bubble is small compared to the horizon.  Furthermore, the bubble interior is initially curvature dominated, so $a(\tau) \approx \tau$, independent of $H$. We can therefore extend the false vacuum
spacetime outside the bubble into the bubble interior up to some small time $\tau_{\epsilon}\ll H_F^{-1}$. The string shape and velocity at $0 < \tau_i < \tau_{\epsilon}$, namely  $\xi \left(\tau_i,\phi\right)$ and $\dot{\xi}\left(\tau_i,\phi\right)$, can then be found from Eq.(\ref{loop1}).  Hence, as $\tau_i \to 0$, $\xi (\tau_i,\phi)$ should satisfy
\ba\label{intcond}
A\left(\eta_n, \chi_n, H_F\right)\tau_i\cosh \xi - B\left(\eta_n, \chi_n, H_F\right) \tau_i\cos \phi \sinh \xi =1,
\ea
where we have defined
\ba
&&A\left(\eta_n, \chi_n, H\right) = H\frac{\cos \eta_n}{\cos \chi_n},\nonumber\\
&&B\left(\eta_n, \chi_n, H\right)= H\tan \chi_n.
\ea
With these initial conditions, the following evolution inside the bubble can be found by solving numerically the equation of motion of the string.

The solution can also be found analytically as follows. We already know that for a string moving in a pure de Sitter space with $H=H_T$, there is a class of solutions which satisfy
\ba\label{solT}
\cos \eta_n' \cosh \xi \tanh H_T \tau - \sin \chi_n' \cos \phi \sinh \xi \tanh H_T\tau -\cos \chi_n' =0.
\ea
Eq.~(\ref{solT}) is the same as Eq.~(\ref{loop2}), except for a difference in $H$. We then observe that the solution described by Eq.~(\ref{solT}) will have the same shape and velocity as the string described by Eq.~(\ref{intcond}) in the limit of $\tau \to 0$, as long as
 \ba\label{AABB}
&&A\left(\eta_n, \chi_n, H_F\right) = A\left(\eta_n', \chi_n', H_T\right), \nonumber \\
&&B\left(\eta_n, \chi_n, H_F\right) = B\left(\eta_n', \chi_n', H_T\right).
\ea
In other words, for a loop nucleated at $\eta=\eta_n$ and $\chi = \chi_n$ in the false vacuum, the evolution of its part inside the bubble can be described by Eq.~(\ref{solT}), with $\eta_n'$ and $\chi_n'$ found through Eq.~(\ref{AABB}). This gives
\ba\label{apcp}
\frac{\cos \eta_n}{\cos \chi_n}=\gamma \frac{\cos \eta_n'}{\cos \chi_n'}~~~\tan \chi_n =\gamma\tan \chi'_n,
\ea
where $\gamma \equiv H_T/H_F$. We also express this mapping in terms of $t_n$ and $r_n$ for later use:
\ba\label{apcp2}
&&r_n'= \frac{2 r_n c^2}{(1-\gamma)r_n^2c^2+(1+\gamma)c^2+\gamma-1}, \nonumber\\
&&c'=\frac{(1-\gamma )c^2r_n^2+c^2 (1+\gamma )-1+\gamma}{\sqrt{\left(\gamma ^2-1\right)c^4\left(r_n^2-1\right)^2+2\left(1-\gamma ^2\right)c^2r_n^2+2\left(1+\gamma ^2\right)c^2-\left(1-\gamma ^2\right)}},
\ea
where we have defined $c=e^{H_Ft_n}$ and $c'=e^{H_Tt_n'}$.  For $\gamma=1$, we get back to the case of $\eta_n'=\eta_n$ and $\chi_n'=\chi_n$. For $\gamma \rightarrow 0$, $\chi_n'$ approaches $\pi/2$.

\begin{figure}
  \includegraphics[width=0.5\linewidth]{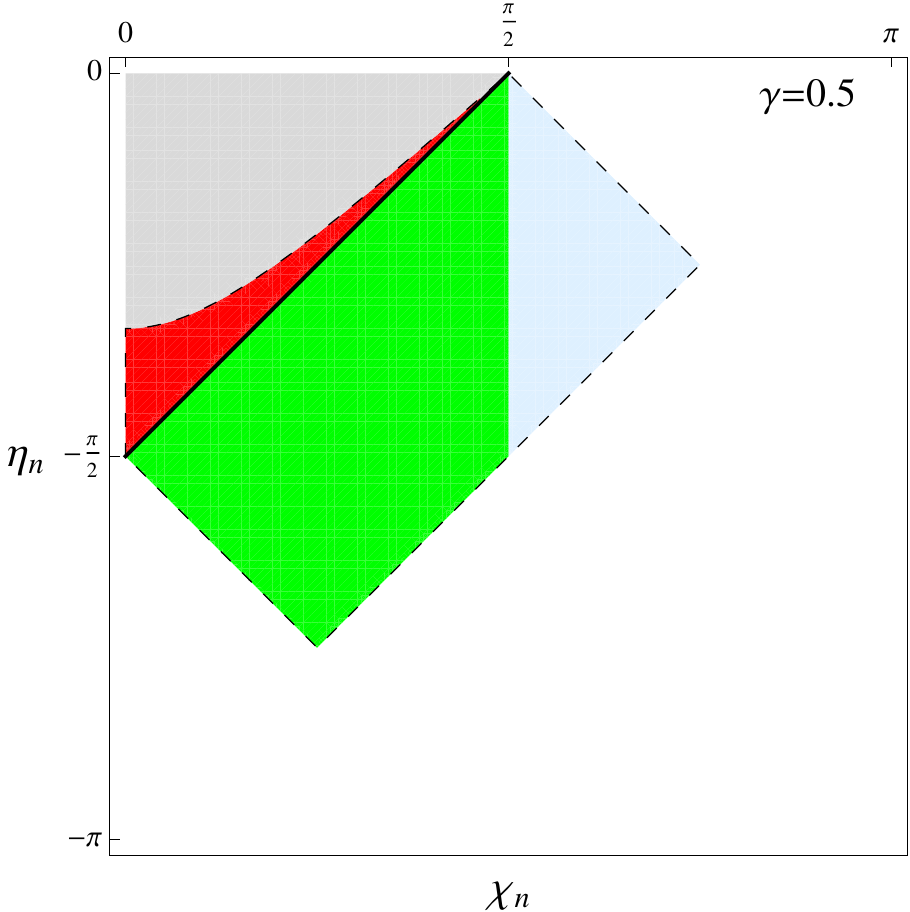}%
  \includegraphics[width=0.5\linewidth]{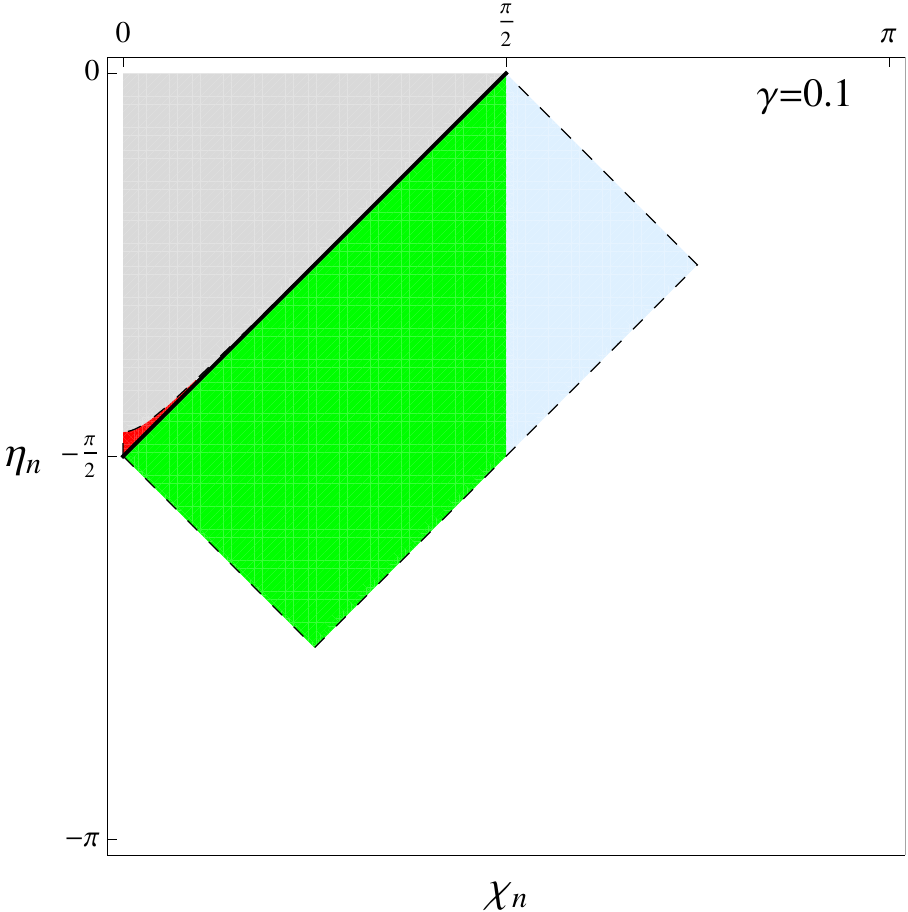}%
\caption{The parameter space of $(\eta_n, \chi_n)$ with $\gamma=~0.5,~0.1$. The color coding is the same as in Fig.~\ref{fig:para}, except that we added a gray color, marking the nucleation centers of loops which collapse after they enter the bubble.}
\label{fig:shift}
\end{figure}

\begin{figure}
  \includegraphics[width=0.8\linewidth]{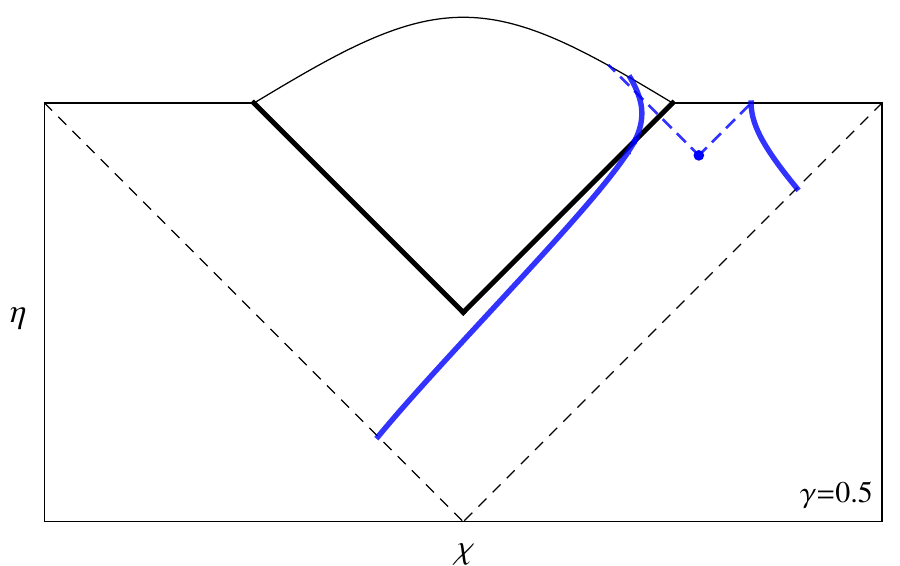}%
\caption{ A string worldsheet with the same nucleation center as that shown in Fig.~\ref{fig:cfdb} in the case when $\gamma = 1/2$.}
\label{fig:cdfb2}
\end{figure}
\begin{figure}
  \includegraphics[width=0.8\linewidth]{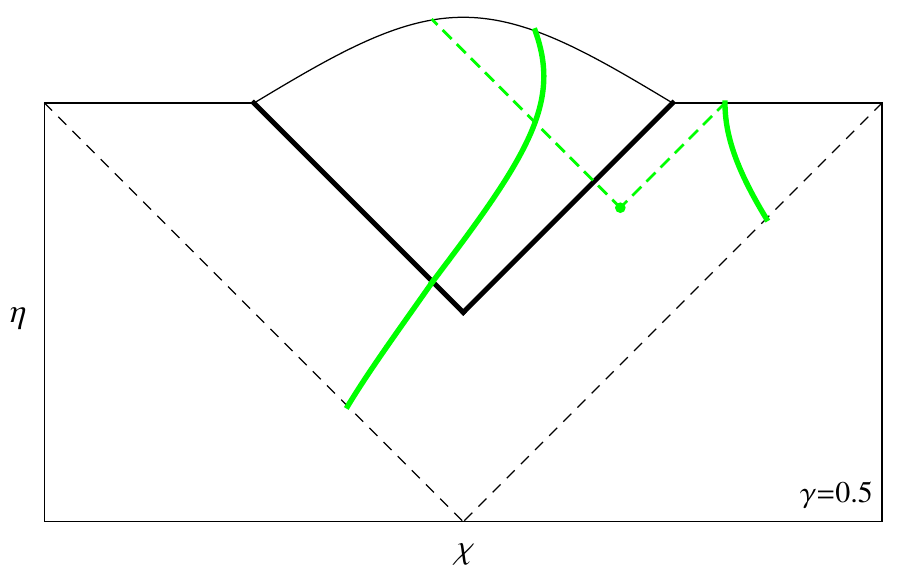}%
 \caption{A string worldsheet with the same nucleation center as that in Fig.~\ref{fig:cfdg} in the case when $\gamma = 1/2$. }
 \label{fig:cdfg2}
\end{figure}
\begin{figure}
  \includegraphics[width=0.8\linewidth]{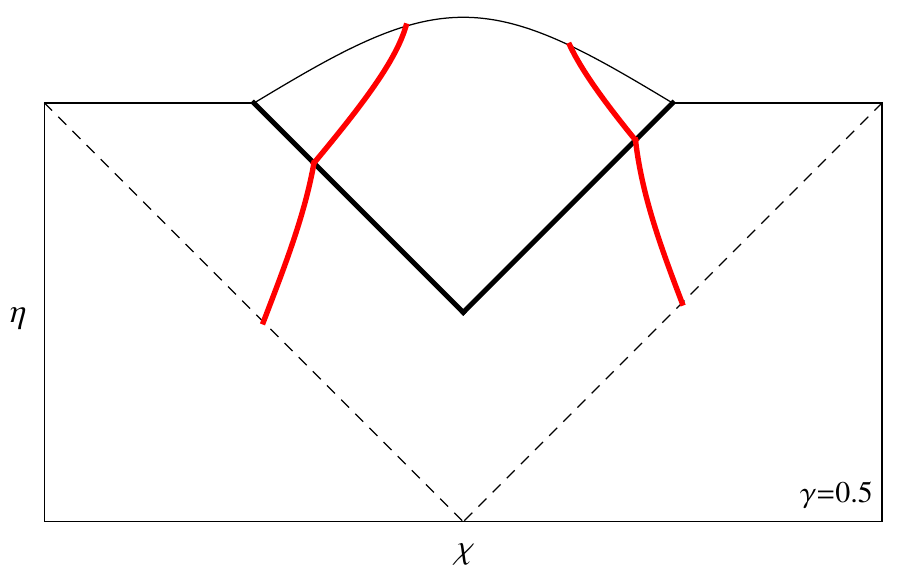}%
\caption{A string worldsheet with the same nucleation center as that in Fig.~\ref{fig:cfdr} in the case when $\gamma = 1/2$. We do not show the nucleation center in this case, because it is in the fictitious false-vacuum de Sitter space.}
\label{fig:cdfr2}
\end{figure}

We have verified that this mapping between solutions agrees well with a numerical calculation. We note, however, that we cannot find a corresponding $(\eta_n',\chi_n')$ for any $(\eta_n,\chi_n)$. For example, for $\chi_n=0$ we have $\chi_n'=0$, and Eq.~(\ref{apcp}) gives $\cos \eta_n = \gamma \cos \eta_n'$. This has no solutions for $\eta_n'$ when $\cos \eta_n > \gamma$. More generally, it can be seen from Eq.~(\ref{AABB}) that, for $(\eta_n,\chi_n)$ such that
\ba
\cos \eta_n > \sqrt{\gamma^2+\tan^2 \chi_n} \cos \chi_n,
\label{ineq1}
\ea
there is no corresponding $(\eta_n',\chi_n')$.  Loops with $\cos \eta_n = \sqrt{\gamma^2+\tan^2 \chi_n} \cos \chi_n$ have ${\eta_n}' = 0$; their physical radius shrinks to the horizon size, $H_T^{-1}$, at $\tau \to \infty$.  And loops satisfying (\ref{ineq1}) collapse to a point in a finite time.  Such loops enter the bubble completely with radii smaller than $H_T^{-1}$ and typically collapse on a timescale $\lesssim H_T^{-1}$ -- unless the inequality (\ref{ineq1}) is only marginally satisfied. The parameter values corresponding to loops collapsing after they enter the bubble are marked by gray color in Fig.~\ref{fig:shift}. Comparing the diagrams with $\gamma=0.5$ and $\gamma=0.1$, we see that, as $\gamma$ decreases, more and more red regions turn into gray. This is because more and more incoming string loops tend to collapse as the horizon inside the bubble increases. We have verified that for the rest of the parameter values $(\eta_n,\chi_n)$ that lead to a collision, one can always find the corresponding $({\eta_n}',{\chi_n}')$.

In Figs.~\ref{fig:cdfb2}-\ref{fig:cdfr2}, we show the worldsheets of strings with the same nucleation centers as in Figs.~\ref{fig:cfdb}-\ref{fig:cfdr} continued into a bubble with $\gamma=1/2$.  Note that the slope of the string worldsheet appears to be discontinuous at the bubble crossing.  This is a spurious effect, which is due to a discontinuous derivative (in the direction normal to the wall) of the conformal factor that we used to generate the conformal diagram.

We note finally that the mapping prescription (\ref{apcp}) was derived only for the case of $\alpha=0$.  We were not able to generalize it for a nonzero $\alpha$.  This method can therefore always be applied to domain walls, but propagation of strings with $\alpha\neq 0$ can only be studied numerically.

\section{Distribution of defects}

Having in mind possible observational effects, we shall now discuss the statistical distribution of strings and walls as it appears to an observer inside the bubble. The potentially observable effects can be roughly divided into two kinds.  If the defect is inside our Hubble radius, it can be observed directly, for example through gravitational lensing effect. In this case, one may be interested in the spatial distribution of strings or walls at a given time $\tau$.  Another possibility is that the defect is no longer within our Hubble radius, but we are in the future light cone of its collision event. We may then find some observational signals of the collision, for example, a distortion of the spectrum of density perturbations, or a burst of gravitational waves emitted during the collision. Then the number of accessible events is just the number of collisions within the past light cone of the observer.

\subsection{Infinite domain walls}

From the point of view of an interior observer, string loops and domain walls that partially enter the bubble can be characterized by the parameters $\xi_m$ and $\Delta\phi$.  In the case of strings, there is also a third parameter, $\alpha$, which is the angle between the plane of the loop and the line connecting the loop and bubble nucleation centers. We shall first consider the simpler case of domain walls. Since the walls get frozen in comoving coordinates shortly after they enter the bubble, their distribution at a late time $\tau$, e.g.,  at the end of inflation, is well approximated by the distribution at $\tau \to \infty$.

\begin{figure}
 \includegraphics[width=0.5\linewidth]{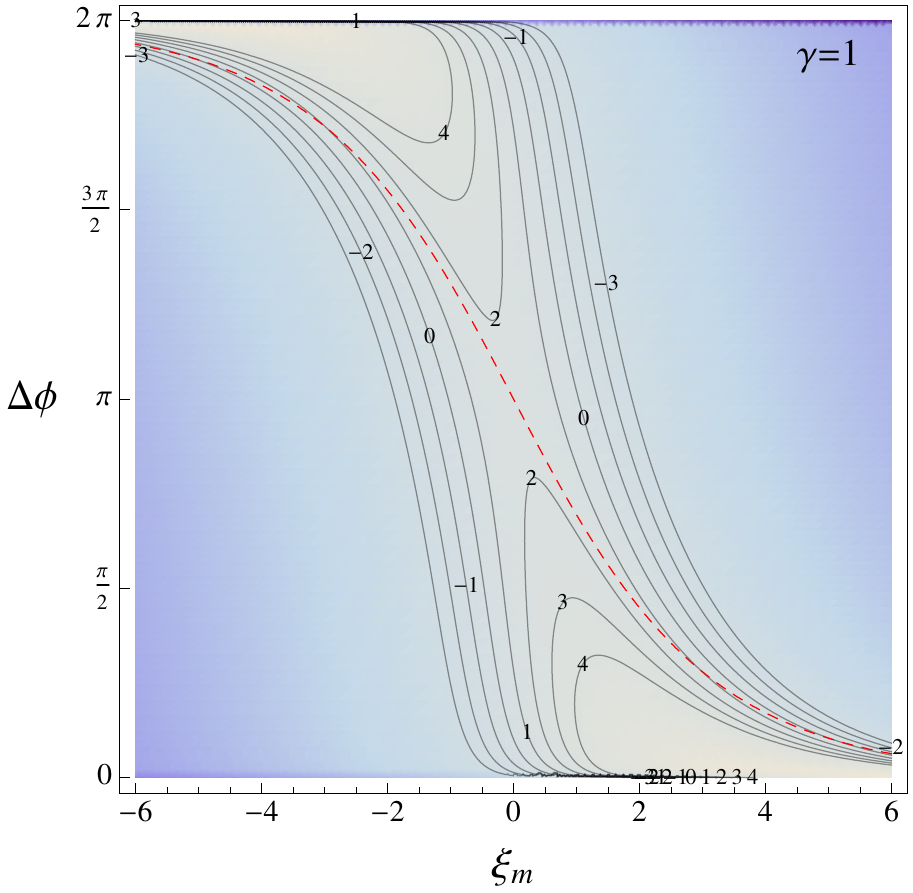}%
  \includegraphics[width=0.5\linewidth]{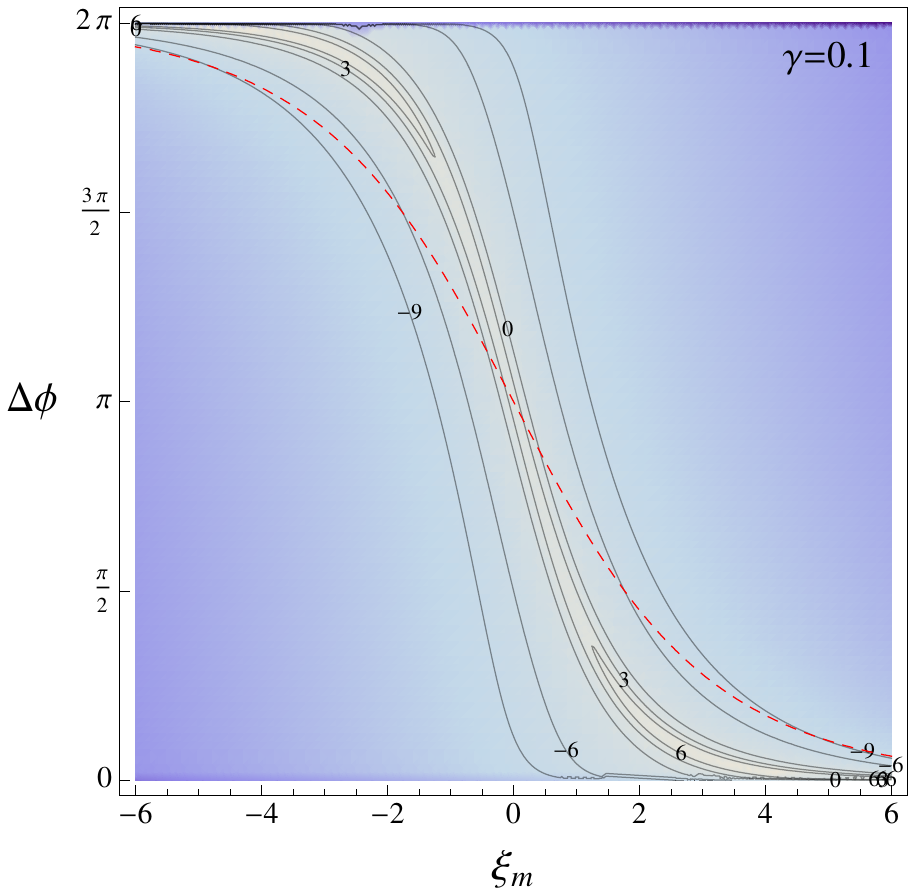}%
\caption{The distribution of $\xi_m$ and $\Delta\phi$ for infinite strings with two different values of $\gamma$. The distribution is only defined in the region below the red dashed line. The numbers labeled on the contours are $\log \frac{dN/\lambda}{d\xi_m d\Delta\phi}$.}
\label{fig:dis}
\end{figure}

We start with the distribution of domain walls with nucleation centers at $(t_n, r_n)$. Then, using Eqs.~(\ref{xim}) and (\ref{dphi}) with $H=H_T$, we can find the distribution with respect to $(\xi_m, \Delta\phi)$.  The calculation is somewhat tedious but straightforward.  The number of domain walls with nucleation centers in spacetime volume element 
\ba
d\Omega_n = e^{3H_F t_n}4\pi r_n^2 dr_n dt_n
\ea
is
\ba 
dN = \Gamma d\Omega_n ,
\ea
where $\Gamma$ is the nucleation rate per unit spacetime volume.  Then we have
\ba
dN &&= \Gamma H_F^{-3} e^{3H_Ft_n} 4\pi r_n^2 dr_n dt_n \nonumber \\
&&={4\pi\lambda} c^2 r_n^2 d r_n dc \\
&&={4\pi\lambda} c^2 r_n^2\left|\frac{\partial\left(r_n,c\right)}{\partial\left(\xi_m,\Delta\phi\right)}\right| d\xi_m d\Delta\phi ,
\label{crn}
\ea
where, as before we have defined
\ba
\lambda\equiv \Gamma H_F^{-4}.
\ea
Using the Jacobian calculated in Appendix \ref{app:JM}, we finally obtain
\ba\label{dis}
\frac{dN}{d\xi_md\Delta\phi} = {32\pi\lambda}\frac{\gamma ^4 \left(2\cosh\xi_m \sin\frac{\Delta\phi }{2}- \sin\Delta\phi\sinh\xi_m \right)}{\left[\left(2\cos \frac{\Delta\phi }{2} \cosh\xi_m-2\sinh\xi_m\right)^2+2\gamma ^2(1-\cos\Delta\phi)\right]^{5/2}}.\ea
This distribution is plotted in Fig.~\ref{fig:dis}. It is defined only in the region
\ba
\frac{\gamma \cos \frac{\Delta\phi}{2}+\sqrt{1-(1-\gamma^2)\cos^2\frac{\Delta\phi}{2}}}{1-\cos\frac{\Delta\phi}{2}}>e^{\xi_m},
\ea
which is below the dashed red line in the figure. This is because the flat chart of de Sitter space covers only the region of $\chi_n - \eta_n \le \pi$. 

At a first glance, Eq.~(\ref{dis}) may suggest that the number of defects is strongly suppressed by a factor of $\gamma^4$ for small $\gamma$. However, as can be seen in Fig. \ref{fig:dis}, there is a band near $\tanh\xi_m=\cos\Delta\phi/2$ where the distribution is enhanced. In the vicinity of this band,
\ba
\frac{dN}{d\xi_md\Delta\phi} \approx \frac{2 \pi\lambda}{\gamma }\cosh^3\xi_m .
\ea

A useful characteristic of the distribution is the total number of defects $N(\xi)$ in the region within a distance $\xi$ from the origin.  This is given by
\ba\label{N1}
N\left(\xi\right)=\frac12\int_{-\xi}^{\xi}d\xi_m \int_{0}^{2\pi}d\Delta\phi \frac{dN}{d\xi_md\Delta\phi}= \frac{4\pi\lambda}{3} \left[2 \xi+\left(1+\gamma^2\right) \sinh2\xi\right].
\ea
Here, the factor $1/2$ accounts for the fact that the distribution (\ref{dis}) is defined only in a half of the full parameter space $(\Delta\phi,\xi_m)$.

Observational bounds on the curvature of the universe indicate that our present Hubble radius cannot exceed 10\% of the curvature radius. This means that a central observer can detect only defects located at $\xi<\xi_h \lesssim 0.1$.  The number of domain walls; within the observable region should then satisfy
\ba
N\left(\xi_h\right) \approx 
\frac{16\pi\lambda}{3}\xi_h \left(1+\frac{\gamma^2}{2}\right) \lesssim 1.7 \lambda .
\label{Ndef}
\ea
With $\lambda \lesssim 1$, we thus expect no more than ${\cal O}(1)$ walls in the observable universe. 
This number may be somewhat enhanced if there is a large number of different kinds of domain walls, as in the axiverse picture \cite{axiverse}.

We note finally that the analysis in this section does not account for the constraint that the defects should be completely outside the bubble at the time of their formation.  Some of the domain walls included in the distributions (\ref{N1}) and (\ref{Ndef}) do not satisfy this constraint, and thus these distributions overestimate the number of defects.  We expect, however, that they still give the right order of magnitude estimates.  (We have verified that this is indeed the case for $\gamma \ll 1$.)

\subsection{Infinite strings}

The calculation of domain wall distribution in the preceding subsection cannot be extended to the case of strings because, as we mentioned at the end of Sec.~III.B, the mapping prescription (\ref{apcp}) can only be used for strings with inclination angle $\alpha=0$.  In order to see the qualitative differences introduced by a variable $\alpha$, we shall restrict the analysis to the case of $\gamma=1$, discussed in Sec.~III.A, where such a mapping is unnecessary.

The distribution of string loops in the parent vacuum is given by
\be
dN = 4\pi \lambda  r_n^2  c^2 ~dr_n~ dc~ d(\sin\alpha),
\label{Nrncalpha}
\ee
where $c = e^{H_F t_n}$, as before.
A loop with a nucleation center at $(t_n,r_n)$ and inclination angle $\alpha$ will overlap with a sphere of radius $r$ centered at the origin if $\alpha\leq \alpha_m(r_n,c,r)$, where 
\be
\alpha_m = \arccos \left[ \frac{c}{2r_n}\left(r_n^2+c^{-2} -r^2\right)\right].
\ee
{The total number of loops within a comoving radius $r$ from the origin is thus given by
\be
N(r) = 4\pi\lambda \int_D dc c^2~dr_n r_n^2 \sin \alpha_m (r_n,c,r),
\label{Nr}
\ee
where the integration domain $D$ is specified by the conditions
\be
c^{-1}-r < r_n < c^{-1} +r,
\ee
\be
r_n>1-c^{-1},
\ee
\ba
c>0.
\ea
}
The first of these conditions requires that a loop with a nucleation center at $(t_n,r_n)$ and $\alpha=0$ overlaps with the sphere at $t\to\infty$.  Some loops with nonzero $\alpha$ will not overlap with the sphere; this is accounted for by the factor $\sin\alpha_m$.  The second condition requires that the loop nucleation center is outside of the bubble.

The integrations in Eq.~(\ref{Nr}) are rather tedious.  {The details are given in Appendix \ref{app:ncs}; }here we only give the final answer:
\be 
N(r) = \frac{\lambda\pi^2 r^2(r^2+16r+4)}{(1-r^2)^2} .
\label{Nr2}
\ee
In the asymptotic region $t\to\infty$, the comoving radius $r$ can be expressed in terms of the $\xi$-coordinate inside the bubble, 
\be
r=\frac{\sinh \xi}{1+\cosh\xi}.
\ee
Substituting this in (\ref{Nr2}), we obtain
\be
N(\xi) = \frac{\lambda\pi^2}{2} \sinh^2 \frac{\xi}{2} (3+5\cosh \xi+16\sinh \xi).
\ee

An important difference of this distribution from Eq.~(\ref{Ndef}) for domain walls is its behavior at small $\xi$,
\be
N(\xi)\approx \lambda\pi^2 \xi^2  ~~~~(\xi\ll 1) .
\ee
The expected number of strings within the observable region is thus
\be
N(\xi_h) \lesssim 0.1 \lambda.
\ee
This is small for any $\lambda\lesssim 1$.

The difference between the wall and string distributions at small $\xi$ is due to the fact that string loops with a sufficiently large inclination angle $\alpha$ do not cross a small sphere near the origin, even if a domain wall of the same radius would cross it.  We expect that the string distribution for $\gamma<1$ will exhibit the same qualitative behavior.

\subsection{Closed strings and walls}

Closed strings and walls, which completely enter the bubble, have their nucleation centers in the bubble interior.  We begin by considering a loop or domain wall whose nucleation center is at $\xi_n=0$. The corresponding worldsheet is given by
\ba
\cosh \xi = \frac{1}{\tanh H_T \tau \tanh H_T \tau_n'},
\label{closed}
\ea
where
\ba
\tanh H_T \tau_n'=\gamma^{-1}\tanh H_F \tau_n.
\ea
The size of the defect can be characterized by the radius
\ba
R=H_T^{-1}\sinh H_T\tau \sinh \xi = H_T^{-1} \cosh H_T \tau \left( \coth^2 H_T{\tau_n}' - \coth^2 H_T\tau \right)^{1/2}  . 
\label{R}
\ea
The worldsheet of a defect with a nucleation center at $(\tau_n,\xi_n)$ can be obtained from (\ref{closed}) by a de Sitter boost (\ref{boost}) with $\beta=\tanh \xi_n$.  
The resulting loop or domain wall will look, respectively, like a circle or a sphere of radius (\ref{R}) from the new nucleation point.

As in the preceding subsection, we shall disregard the constraint that defects can only form outside of the bubble.  Our result will therefore be an overestimate of the density of closed defects.  We shall see, however, that this estimate still tends to be rather small.

The number of defects with nucleation centers in the interval $(d\tau_n,d\xi)$ is given by
\ba
dN= \Gamma d\Omega_4 = 4\pi\lambda H_F \sinh^2\xi d\xi \sinh^3 H_F \tau_n d \tau_n.
\ea
Note that the 4-volume element $d\Omega_4$ is in a fictitious de Sitter space, obtained by continuing the parent false vacuum into the bubble interior.  The nucleation centers are uniformly distributed in this extended region (with a constant density per unit 4-volume).  This distribution is invariant under de Sitter boosts (\ref{boost}), which implies that the defect distribution should have the same symmetry.  With these assumptions, the defects are uniformly distributed on the surfaces of constant time $\tau$ inside the bubble.\footnote{The cutoff at $t_f$ would amount to requiring that defects can only be formed with $t_f < 0$, or equivalently with $t_n \lesssim S_E $.   This cutoff would of course break the de Sitter invariance.   
We expect that it would suppress the closed defect density at large values of $\xi$.  The effect of the cutoff at $t_f$ on the number of observable collisions will be discussed in Section V.}

To find the distribution of defects of a given radius $R$ on a hypersurface of constant $\tau$, we express $\tau_n$ in terms of $R$ from Eq.~(\ref{R}), 
\ba\label{disl}
dN=  \frac{\lambda \gamma^4 dV }{\tanh^4H_T\tau\sqrt{R^2+R_0^2}}\frac{R_0RdR}{\left(R^2+R_{0}^2-\gamma^2(H_T^{-2}+R_{0}^2)\right)^{5/2}}.
\ea
Here, $dV=4\pi H_T^{-3}\sinh^3H_T\tau\sinh^2\xi d\xi$ is the volume element on the hypersurface, and $R_0(\tau)=H_T^{-1}\sinh H_T \tau$ is the curvature radius of the spacelike slice $\tau={\rm const}$.
For $\gamma \ll 1$ and $H_T \tau \gg 1$, the distribution simplifies to
\ba
dN \simeq \lambda \gamma^4 dV \frac{R_0 R dR}{\left(R^2+R_0^2\right)^3}.
\ea
We can also define the comoving size as $\widetilde{R} = R / R_0$ and the comoving volume $d\widetilde{V} = dV \sinh^{-3}H_T\tau$.  Then 
\ba
dN  \simeq  \lambda\gamma^4 d\widetilde{V} \frac{\widetilde{R}d\widetilde{R}}{(\widetilde{R}^2+1)^3},
\ea
which simply says that the defects are formed at early times $\tau$ and are then stretched and diluted by the expansion, leaving the comoving distribution unchanged. 

The distribution (\ref{disl}) shows that the number of defects is strongly suppressed by $\gamma^4$. This suppression is in qualitative agreement with the shrinking of the red region in Fig.~\ref{fig:shift}. Hence, we should not expect to see any closed loops or walls from the multiverse within our horizon, unless $\gamma\sim 1$.

Closed strings and walls can also nucleate in the true vacuum inside the bubble. The corresponding nucleation rates are given by Eq.~(\ref{Gamma}) with the instanton action (\ref{SE}) or (\ref{SE2}) and with $H_F$ replaced by $H_T$. This rate, however, will be exponentially suppressed for $H_T < H_F$, assuming that the string or wall tension is the same inside and outside the bubble. If the true vacuum admits lighter defects, which do not exist in the false vacuum, they could have a higher nucleation rate and could produce some observable effects (see e.g.~\cite{Garriga:1992nm}). However, observation of such defect would not provide evidence for a multiverse, so we do not consider them here.

\section{The number of collision events}

We shall now estimate the number of collision events within the observer's past light cone.  We consider a central observer, located at $\xi = 0$ inside the bubble.  The past light cone of such an observer, originating at time $\tau_O$, is given by
\ba
e^\xi = \frac{\tanh(H_T \tau_O/2)}{\tanh(H_T\tau/2)}.
\ea
Its physical radius at time $\tau <\tau_O$ is
\ba
R(\tau) = H_T^{-1}\sinh\xi \sinh H_T\tau = H_T^{-1} \coth(H_T\tau_O/2) \left[1-\frac{\cosh^2(H_T\tau/2)} {\cosh^2(H_T\tau_O/2)}\right].
\ea
The bubble wall in these coordinates is at $\tau = 0$; it intersects the past light cone on a sphere ${\cal S}$ of radius
\ba
R(0) = H_T^{-1}\tanh(H_T\tau_O/2).
\ea
For $\tau_O\gg H_T^{-1}$, we have $R(0)\approx H_T^{-1}$ -- which is to be expected, since $H_T^{-1}$ is the horizon radius in the de Sitter space of the bubble.  

\begin{figure}
 \includegraphics[width=0.8\textwidth]{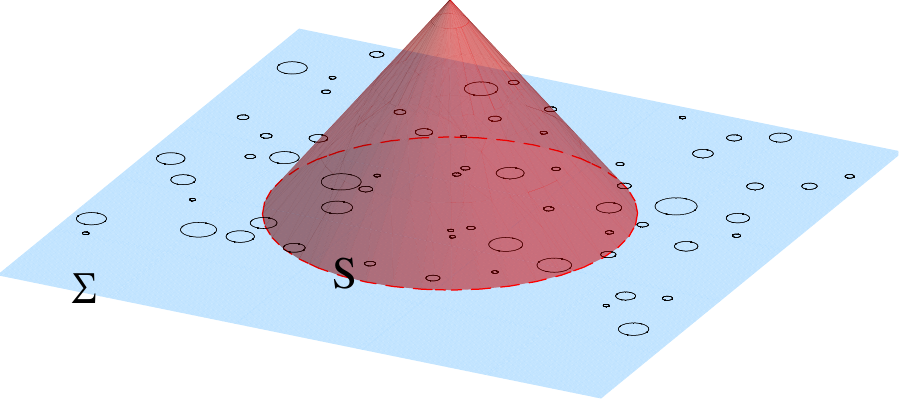}
 \includegraphics[width=0.9\textwidth]{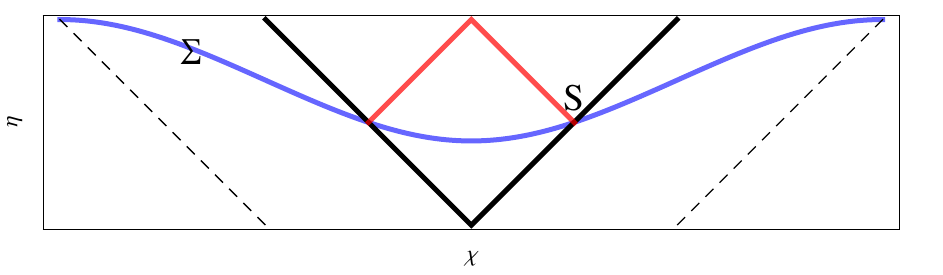}
\caption{The red cone is the past light cone of the observer which intersects the bubble wall on the sphere ${\cal S}$, and the blue plane is the hypersurface $\Sigma$. Black circles show the domain walls on that hypersurface.}
\label{fig:s}
\end{figure}

Let us consider the particular hypersurface of constant time $\Sigma: t=t_l={\rm const}$ in the flat slicing coordinates given by Eq. (\ref{dS}), which includes the sphere ${\cal S}$. From Eq.~(\ref{wallworldsheet}) we have
\be
e^{H_F t_l} = \frac{1+\gamma}{\gamma}.
\label{t0gamma}
\ee
This hypersurface extends from the sphere into the false vacuum {(see Fig.~\ref{fig:s})}. Now, the number of defects of radius $R$, whose centers are located in a 3-volume element $dV$ on any constant-time surface, is given by Eq.~(\ref{dN}), 
\ba
dN = \lambda  \frac{dR}{R^4} dV,
\ea
with a lower cutoff at $R\approx H_F^{-1}$. The defects that intersect the sphere ${\cal S}$ at $t=t_l$ must have collided with the bubble wall at some earlier time $t_c$.  We shall estimate the number of such collisions, focussing first on domain walls.

It will be convenient to use the units in which $H_F=1$ and $H_T=\gamma$.
The collision time $t_c$ for a domain wall with nucleation center at $(t_n,r_n)$ can then be found from
\be
e^{t_c -t_n} = \frac{2z}{1-z^2},
\label{z}
\ee
where $z=(r_n-1)e^{t_n}$.  It should satisfy 
\be
t_f<t_c<t_l,
\label{tftc}
\ee 
where the time of wall formation $t_f$ is related to $t_n$ by Eq.~(\ref{tf}).  With the aid of Eqs.~(\ref{z}), (\ref{tf}), the condition $t_c>t_f$ can be expressed as
\be
e^{t_c-t_f} = e^{t_c-t_n}e^{t_n-t_f} \sim \frac{2z}{1-z^2} S_E > 1,
\ee
or
\be
z\gtrsim \sqrt{S_E^2 +1} -S_E.
\ee
For $S_E\gg 1$, this gives
\be
(r_n-1)e^{t_n} \gtrsim\frac{1}{2S_E}.
\label{rnSE}
\ee
The parameters $t_n$, $r_n$ can be expressed in terms of the domain wall radius $R$ and the physical distance $r_{phy}$ from its center to the bubble center,\footnote{Note that the distance $r_{phy}$ is calculated in the fictitious de Sitter space, obtained by extending the false vacuum to the bubble interior.}
\be
e^{-t_n}=\frac{\gamma}{1+\gamma} \sqrt{R^2-1},
\ee
\be
r_n = e^{-t_l}r_{phy} = \frac{\gamma}{1+\gamma} r_{phy}.
\ee
Substituting this in (\ref{rnSE}) we obtain
\be
r_{phy}>\frac{1}{\gamma} + 1 +\frac{1}{2S_E} \sqrt{R^2 -1}. 
\ee
For the most numerous domain walls with radii near the lower cutoff, $R\ll S_E$ and the above condition simplifies to
\be
r_{phy}>\frac{1}{\gamma} + 1. 
\label{cond1}
\ee

Similarly, requiring that $t_c<t_l$, we obtain the condition
\be
r_{phy} < \frac{1}{\gamma}+R.
\label{cond2}
\ee
It simply states that the domain wall should be close enough to overlap with the bubble.

The total number of intersections can now be found by counting all domain walls in the distribution (\ref{dN}) that satisfy the conditions (\ref{cond1}), (\ref{cond2}),
\ba
{\cal N} \gtrsim \lambda  \int_{1}^{\sim \gamma^{-1}} \frac{dR}{R^4} \int_{\gamma^{-1}+1}^{\gamma^{-1}+R} 4\pi r_{\text{phy}}^2 dr_{\text{phy}}
\approx \frac{2\pi}{3}\lambda \gamma^{-2} ,
\label{calN}
\ea
where in the last step we assumed that $\gamma\ll 1$. We see that the number of potentially observable collisions is enhanced by a factor of $\gamma^{-2}$ compared to the number of walls (\ref{Ndef}) that are physically present within the visible universe.  A similar enhancement factor was found in Ref.~\cite{Freivogel} for observable bubble collisions.  For $\gamma\ll 1$ it is possible to have ${\cal N}\gg 1$, provided that the wall nucleation rate $\lambda$ is not very strongly suppressed.

The total number of intersections for cosmic strings ${\cal N}_s$ can be found by a calculation similar to that in Sec. IV.B.  One obtains an integral similar to Eq.~(\ref{calN}), but with a weighting factor $\sin \alpha_m(R, r_{\text{phy}},\gamma)$, due to the variable inclination angle. $\alpha_m$ is the maximal inclination angle for given $R$ and $r_{\text{phy}}$,
\ba
\alpha_m = \arccos\left[\frac{r_{\text{phy}}^2+R^2-\gamma^{-2}}{2r_{\text{phy}} R}\right].
\ea
For small $\gamma$, we find
\ba
{\cal N}_{s} \approx \frac{15\pi^2}{16}\lambda \gamma^{-2}.
\label{calNcs}
\ea

In the above analysis we disregarded defects that have their nucleation centers inside the bubble. Such defects encircle the bubble at $t=0$, and we shall assume as before that they can only be formed at $t_f<0$.  Then their nucleation centers must be at $t_n<t_{*}\sim \ln S_E$.  The number of such defects can be estimated as
\be
{\cal N}' \sim \Gamma \Omega_4,
\ee
where $\Omega_4$ is the 4-volume of the relevant spacetime region,
\be
\Omega_4 =  \int_0^{r_*} dr\cdot 4\pi r^2 \int_{t_b(r)}^{t_{*}} dt e^{3t} .
\ee
Here, 
\be
t_b(r) = - \ln (1-r)
\ee
is the worldsheet of the bubble wall and
\be
r_* \sim 1 - \frac{1}{S_E}
\ee
is the comoving radius of the bubble wall at time $t_*$.
Performing the integrations and assuming that $S_E\gg 1$, we obtain
\be
{\cal N}' \sim \frac{2\pi}{9} \lambda S_E^3.
\label{Nenclosed}
\ee
The factor $S_E^3 \gg 1$ appears to enhance the number of collisions; however, the nucleation rate $\lambda\sim e^{-S_E}$ is suppressed at large $S_E$, so we always have ${\cal N}' \lesssim 1$.

\section{Collision marks in the sky}

We shall now try to see what the collision events might look like in the sky.  We shall not attempt any realistic analysis of observational effects that the collisions may produce in the CMB.  Instead, we shall assume that collisions with defects leave some sort of marks, or `scars' at very early times in the bubble $(\tau\to 0)$.  One can assume, for example, that defects disintegrate upon colliding with the bubble and their energy is deposited in close vicinity of the impact.  Assuming that this energy excess somehow becomes observable, we shall be interested in the shape of the resulting `scars' and in their distribution in the observer's sky.

\begin{figure}
  \includegraphics[width=0.5\linewidth]{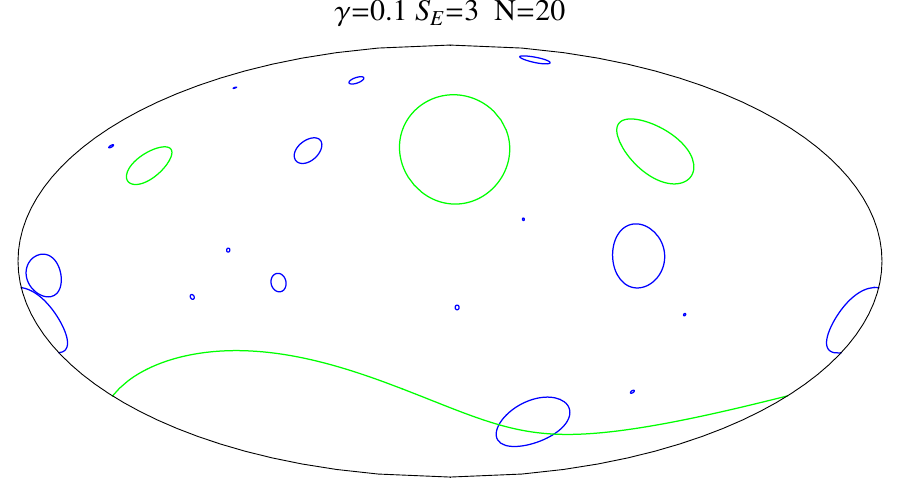}%
  \includegraphics[width=0.5\linewidth]{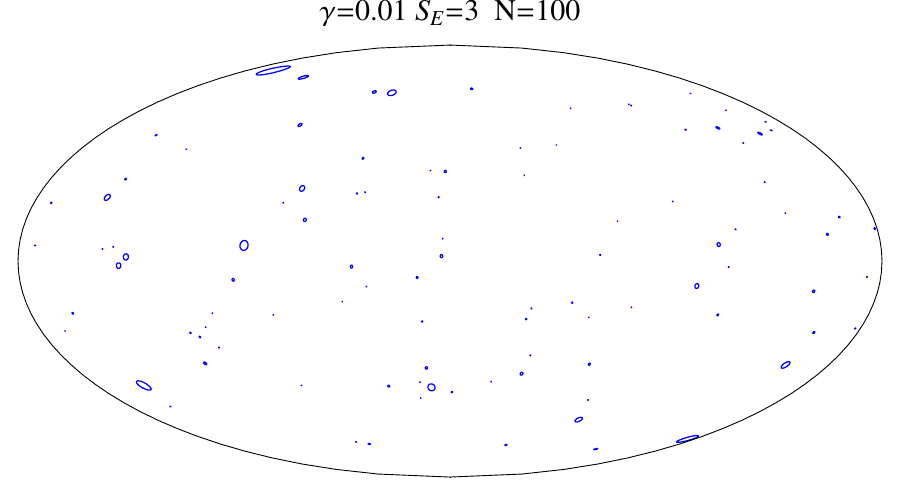}%
\caption{Collision marks of domain walls on the sky. The left panel shows 40 collision events with $\gamma=0.1$, and the right panel shows 100 collisions with $\gamma=0.01$.  The corresponding nucleation rates are $\lambda=0.025$ and $\lambda=0.0016$, respectively.}
\label{fig:dwonsky}
\end{figure}

\begin{figure}
  \includegraphics[width=0.5\linewidth]{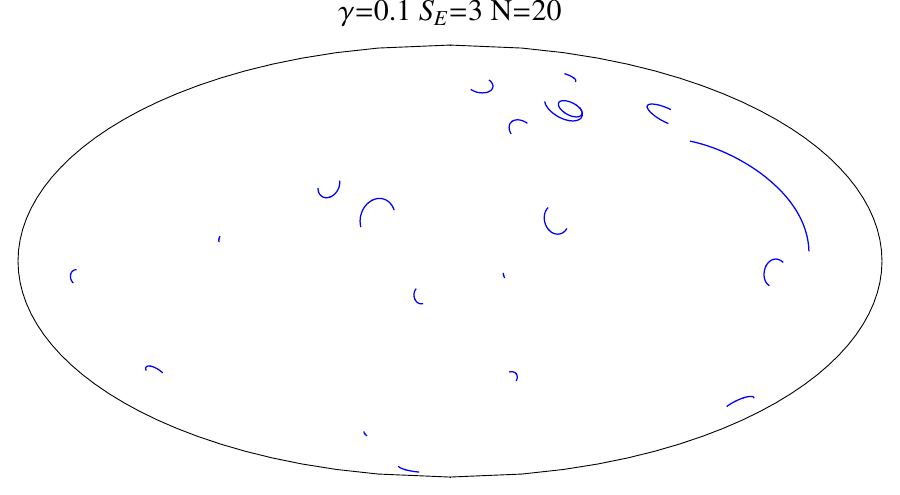}%
  \includegraphics[width=0.5\linewidth]{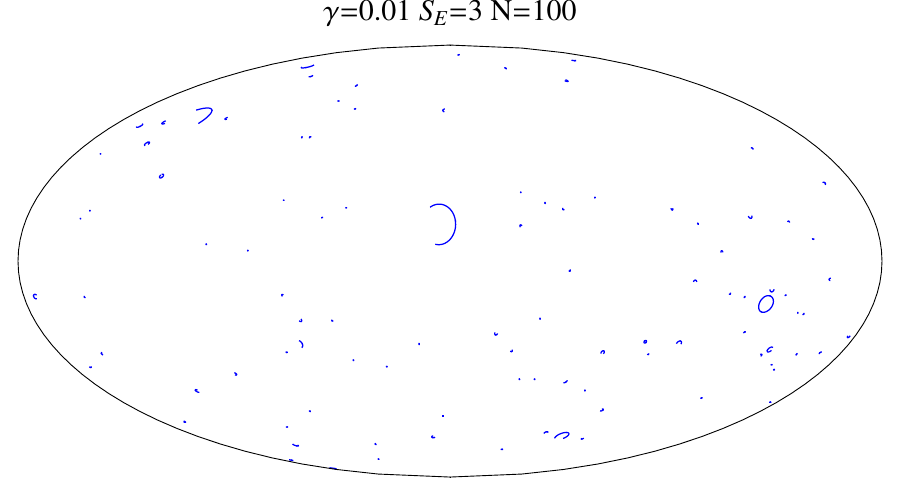}%
\caption{Collision marks of cosmic strings on the sky. The numbers of collision events, the values of $\gamma$ and the color code are the same as in Fig.~\ref{fig:dwonsky}. Note that all strings in these simulations nucleate in the blue region of the parameter space.}
\label{fig:csonsky}
\end{figure}

The mark left by a collision with a domain wall has the form of a disc.  In Fig.~\ref{fig:dwonsky} we show the distribution of wall collision marks in the sky for two values of the parameter $\gamma$, using the Mollweide projection of the celestial sphere.  Note that this projection distorts the circular shape of the marks.  The distributions were generated for a fixed number of collisions ${\cal N}$, with a random choice of nucleation centers.  We only included walls that are completely outside the bubble at the time of their formation $t_f$.  Blue and green curves in the figure are respectively the boundaries of collision marks due to walls with nucleation centers in the blue and green areas of the parameter space of Fig.~\ref{fig:shift}.  The green collision marks correspond to domain walls that formed prior to the bubble formation and initially enclosed the bubble.  A typical bubble can be expected to collide with such walls only if the wall nucleation rate is relatively high, as in the left panel of Fig.~\ref{fig:dwonsky}.
Blue collision marks are due to walls that formed outside the bubble. 

For blue curves the area of impact on the sky is enclosed by the curve; in other words, the affected area is always less than half of the sky.  For green curves the affected area can either be inside or outside, but we find that it is predominantly outside.  For example, all of the green curves in the left panel of Fig.~\ref{fig:dwonsky} have the affected region outside.  

For a head-on $(\alpha=0)$ collision with a string loop, the mark is just a straight segment, while for a collision with $\alpha\neq 0$ the mark segment is curved.  Two simulated distributions of string collision marks in the sky are shown in Fig.~\ref{fig:csonsky}.  As for the walls, the distributions were generated for a fixed number of collisions, but now the nucleation centers and loop inclination angles $\alpha$ had to be chosen at random. The details of the simulations are given in Appendix D.

\begin{figure}
  \includegraphics[width=0.5\linewidth]{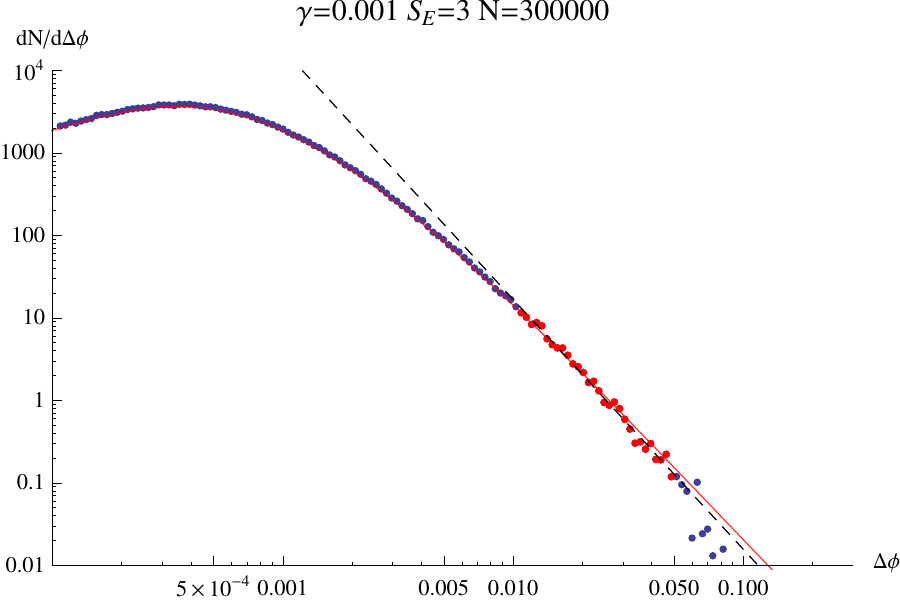}%
  \includegraphics[width=0.5\linewidth]{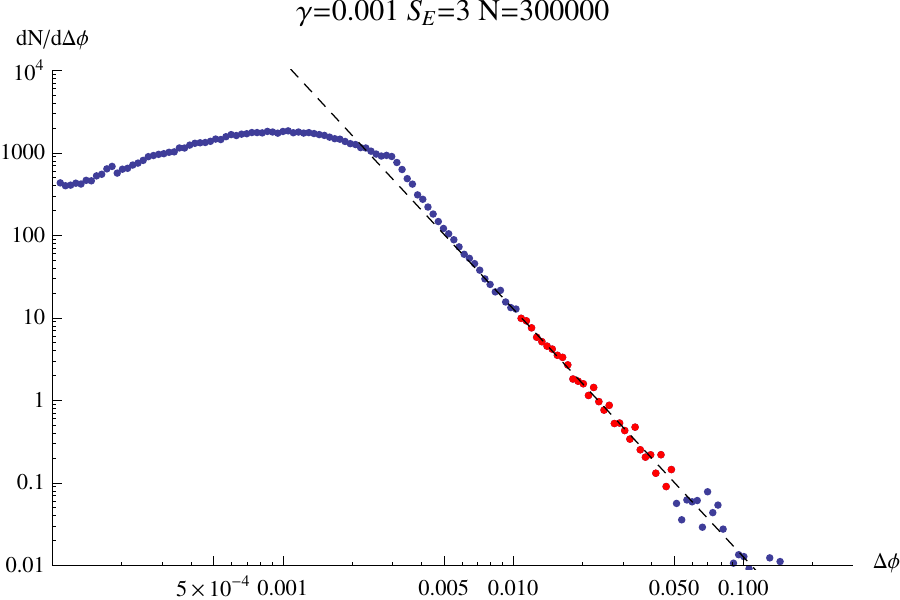}%
\caption{Numerically generated angular size distributions for 300,000 collision marks made by domain walls (the left plot) and by cosmic strings (the right plot). In the left plot, the analytically calculated distribution for domain walls is shown by a solid red line. We fit the distributions with a power law (\ref{fit}) in the range of $0.01< \Delta\phi<0.05$. (The points used in the fits are shown in red.) The best fits (shown by dashed black lines) yield the values $p = -3.03192$ for domain walls and $p=-3.00792$ for strings.}
\label{fig:phidis}
\end{figure}

The distribution of angular sizes of domain wall collision marks on the sky can be found analytically by calculating the 4-volume of the appropriate nucleation region.  The result is shown in Fig.~\ref{fig:phidis} by the red solid line.  The calculation involves some tedious integrations, and we do not reproduce it here.  Instead, we shall find the form of the distribution for relatively large angles, which can be calculated as follows.

As in the preceding section, it will be convenient to use units in which $H_F =1$ and  $H_T = \gamma$.  
Once again, we consider the spherical surface ${\cal S}$, defined by the intersection of the observer's past light cone and the bubble worldsheet, and the flat hypersurface $\Sigma: t=t_l$ that includes ${\cal S}$.  Now consider a domain wall of radius $R\ll \gamma^{-1}$ lying in the hypersurface $\Sigma$ with its center at a distance $\rho<R$ from ${\cal S}$. The intersection of the wall with the surface ${\cal S}$ will be seen from the origin at an angle
\ba
\Delta\phi \approx 2 \gamma \sqrt{R^2 -\rho^2}.
\label{dphic}
\ea
Using the wall distribution (\ref{dN}) and expressing $\rho$ in terms of $\Delta\phi$ from Eq.~(\ref{dphic}), we have
\ba
dN = 2\pi\lambda \gamma^{-4} \frac{dR~\Delta\phi d\Delta\phi}{R^4\sqrt{4R^2 -\gamma^{-2}\Delta\phi^2}} .
\label{dN2}
\ea

To find the distribution of angular sizes $\Delta\phi$, we need to integrate Eq.~(\ref{dN2}) over $R$, with a lower limit of integration at $R_{min}=\Delta\phi /2\gamma$.  This gives
\ba
dN  = \frac{32\pi\lambda}{3} \frac{d\Delta\phi}{\Delta\phi^3} .
\label{phi-3}
\ea
This should be accurate, as long as $\gamma^{-1}\gg R_{min}\gg 1$, that is, $1\gg \Delta\phi \gg 2\gamma$.

Numerically generated angular size distributions for collision marks made by domain walls and cosmic strings are also shown in Fig.~\ref{fig:phidis}. They were obtained by averaging over 300,000 instances of randomly chosen defect nucleation centers and using the parameter values of $\gamma=0.001$ and $S_E=3$.  The vertical axis in the figure shows the quantity $dN/d(\Delta\phi)$ (up to an arbitrary factor).  For domain walls, the numerically generated distribution agrees very well with the analytic result. For cosmic strings, we defined the angular size $\Delta\phi$ as the angle between the end points of the string collision mark on the sky.  In this case, we did not attempt to derive the distribution analytically, so we show only the numerically generated distribution.     

We fitted the distributions with a power law 
\ba
dN \propto (\Delta\phi)^p d(\Delta\phi)
\label{fit}
\ea
in the range\footnote{To reduce noise in the numerical data, we used logarithmic bins for $\Delta\phi$.  The fitting range for the power law was chosen so that the number of collision marks per bin is no less than $18$ for all bins in the range.  The lower bound of the range, $\Delta\phi > 0.01$, was chosen to satisfy the condition of validity of the power law approximation, $\Delta\phi\gg 2\gamma = 0.002$.}  of $0.01< \Delta\phi<0.05$.  The best fit yields the values $p = -3.03192$ for domain walls and $p=-3.00792$ for strings, suggesting that the string distribution is likely to approach the same asymptotic form (\ref{phi-3}).

The analysis of bubble collisions in Ref.~\cite{GGV} has shown that the distribution of collision events on observer's sky is generally anisotropic.  Assuming that inflation begins on a surface $t=t_i$ in the flat slicing chart (\ref{dS}), only the central observer, who is comoving with respect to the geodesic congruence orthogonal to this initial surface will see an isotropic distribution.  This effect, which persists even in the limit $t_i\to\infty$, has been dubbed ``the persistence of memory".  If the coordinates inside the bubble are chosen so that the central observer is at the origin, $\xi=0$, then for all other observers the distribution will be peaked in the direction away from the origin.  We can, of course, perform a de Sitter boost that brings any given observer to $\xi=0$, but this transformation will also affect the initial surface, which will now slope down in time in the direction opposite to the boost.  The number of bubble collisions, which is proportional to the spacetime volume in the past light cone of the observer with a cutoff at the initial surface, will peak in the direction of the slope.

It has been shown in \cite{Freivogel} that this persistence of memory effect disappears in the limit of $\gamma \ll 1$. The reason is that the number of collisions in this case is dominated by collisions with the smallest defects of size $\sim H_F^{-1} \ll H_T^{-1}$, whose nucleation centers are in the future of the collision time. The same applies to collisions with nucleating defects.  This is demonstrated by an explicit calculation in Appendix \ref{app:persistence}.

\section{Discussion}

We investigated various aspects of collisions between strings and domain walls nucleating in an inflating false vacuum and our bubble, which we assume to be expanding in that vacuum.  These collisions have some similarity to collisions with other bubbles, but in some respects they are very different.

The character of collision between the bubbles crucially depends on whether the vacuum energy density of the other bubble is higher or lower than that in our bubble.  If it is lower, then the invading bubble will expand into our bubble at nearly the speed of light, and all objects it encounters on its way, including observers, will be turned into some alien forms of matter.  If it is higher, then our bubble will expand into the other bubble, and only a peripheral region of our bubble will be affected by the collision.

In the case of defects, the collisions are always rather benign.  Defects that do not collapse shortly after they enter the bubble, eventually come to rest relative to the comoving observers and can be detected through their gravitational effects.  We found, however, that there can be no more than a few defects from the multiverse in our observable region, and only if their nucleation rate is very high, $\lambda\sim 1$.   Because of their small number, such defects are not subject to observational bounds that have been derived for ``regular" defects, formed in cosmological phase transitions.
The tension $\mu$ of ``regular" cosmic strings is bounded by \cite{BOS} $G\mu \lesssim 10^{-8}$.  Strings from the multiverse, on the other hand, could have, for example, $G\mu\gtrsim 10^{-5}$ without conflict with observational data, provided that the angular extent of these strings on the sky is sufficiently small.  ``Regular" domain walls are essentially ruled out by observations, unless their tension is extremely small, $\sigma\lesssim (1 ~\text{MeV})^3$ \cite{VSbook}.  A wall with a larger tension crossing our observable region would introduce a very large gravitational perturbation, which would be in conflict with the isotropy of the CMB and of the galaxy distribution.  The tension of walls from the multiverse can be greater by many orders of magnitude, if these walls are located at sufficiently high redshifts. The effect of massive defects on the CMB power spectrum has been discussed in \cite{Wise} for a straight cosmic string and in \cite{Jazayeri} for a planar domain wall.  In particular, it has been argued in \cite{Jazayeri} that a massive domain wall at about twice the horizon distance from us could account for the hemispherical power asymmetry indicated by recent observations.

Collisions of walls and strings with our bubble can have observational signatures even if the defects themselves are not within our present Hubble radius.  One can expect the observational effects of collisions with domain walls to be similar to the effects of collisions with other bubbles.  Such collisions will produce round hot or cold spots on the CMB sky and a characteristic CMB polarization pattern \cite{Aguirre1,Kleban1,Freivogel,Aguirre2,Kleban2,Johnson}.  Collisions with strings do not have rotational symmetry and (unlike collisions with domain walls and with other bubbles) can also produce gravitational radiation.  We leave a detailed analysis of observational signatures of collisions with defects for future research.

We have estimated the number of potentially observable collisions with defects in our past light cone as 
\ba
{\cal N} \sim \lambda (H_F/H_T)^2, 
\label{calN2}
\ea
where $H_F$ and $H_T$ are the expansion rates, respectively, of the false vacuum and of the slow-roll inflation inside the bubble.  This number can be large even if the nucleation rate of the defects is $\lambda\ll 1$.

We finally mention some other types of defects which can also nucleate in de Sitter space and collide with our bubble.  In a wide class of particle physics models, domain walls can be bounded by strings \cite{Shafi,Everett}.  In such models, the nucleating defects will have the form of disc-like domain walls bounded by circular loops of string.  The corresponding instanton is the usual domain wall instanton with a hole bounded by a string.  (The same instanton describes nucleation of circular holes in the walls.)   

Monopoles and antimonopoles can be produced in pairs, as discussed in Ref.~\cite{Basu}, with a separation $\approx 2H_F^{-1}$.  Magnetic monopoles behave as localized particles and, considering their small numbers, do not produce any significant effects.  Global monopoles, on the other hand, have a non-local distribution of energy outside of their cores and can create strong gravitational perturbations \cite{Manuel}, both in our Hubble volume and when they enter the bubble in our past light cone.  Global monopole and antimonopole in a pair are attracted to one another with a force independent of their separation, so pairs that enter our bubble withe a separation $< H_T^{-1}$ will collapse and annihilate.  By the same argument as in Sec.~IV.C, the number of potentially observable collision events is still given by Eq.~(\ref{calN2}).

\section*{Acknowledgements}

A.V. is grateful to Matt Kleban for a useful discussion.  This work was supported in part by IKERBASQUE, the Basque Foundation for Science and MINECO under grant number, FPA2012-34456 (JJBP), by MEC FPA2013-46570-C2-2-P, AGAUR 2014-SGR-1474, CPAN CSD2007-00042 Consolider-Ingenio 2010 (JG) and by the National Science Foundation (AV and JZ).  J.Z. was also supported by
the Burlingame Fellowship at Tufts University.

\appendix


\section{Coordinate Systems}
\label{app:coordinates}

In this Appendix we summarize all coordinate systems we used in this paper and the relations between them. The false vacuum spacetime can be embedded in a 5-dimensional Minkowski space with coordinates $({\vec X},W,V)$ as
\ba
\vec{X}^2+W^2-V^2=H_F^{-2}.
\ea
The surface $W=H_F^{-1}$ with $V \ge 0$ corresponds to a bubble wall with nucleation center at ${\vec X}=V=0$.  The spacetime of true vacuum inside the bubble can also be embedded in a 5 dimensional Minkowski space as
\ba
\vec{X}^2+W'^2-V^2=H_T^{-2},
\ea
where $W' \ge H_T^{-1}$ and $V \ge 0$, with the bubble wall at $W' = H_T^{-1}$.

The flat slicing coordinates $\left(t,r,\theta,\phi\right)$ are defined by
\ba\label{flat}
&& V = H_{F}^{-1} \sinh H_F t + \frac{H_F^{-1}}{2}e^{H_F t}r^2 \\ \nonumber
&& X_{i} = H_F^{-1} r e^{H_F t} \omega_i \\ \nonumber
&& W = H_{F}^{-1} \cosh H_F t - \frac{H_F^{-1}}{2}e^{H_F t}r^2,
\ea
with $(\omega_z,\omega_x,\omega_y)=(\cos\theta,\sin\theta\cos\phi,\sin\theta\sin\phi)$. And the open slicing coordinates $\left(\tau,\xi,\theta,\phi\right)$ inside the bubble are defined by
\ba
&& V = H_{T}^{-1}\sinh H_T \tau \cosh \xi  \\ \nonumber
&& X_{i} =H_{T}^{-1}\sinh H_T \tau \sinh \xi  \omega_{i}  \\ \nonumber
&& W' =H_{T}^{-1}\cosh H_T \tau,
\ea
In general there is no transformation between $\left(t,r,\theta,\phi\right)$ and $\left(\tau,\xi,\theta,\phi\right)$, since they cover different regions of spacetime. However, in the case of $H_T=H_F=H$, these coordinate systems are related in the overlap region through
\ba
&&\cosh H\tau=\cosh Ht - \frac12 e^{Ht}r^2 \nonumber \\
&&\sinh H\tau \sinh \xi = e^{Ht} r.
\ea
To represent the whole de Sitter space-time, we use the standard conformal diagram that suppresses a 2-sphere at each 
point and makes use of the conformal version of the closed slicing of de Sitter space, namely,
\ba
ds^2 = {1\over {H^2 \sin^2\eta}} \left(d\eta^2 - d\chi^2 + \sin^2\chi~d\Omega_2^2\right)~.
\ea
The nice thing about this slicing is that it covers the whole space-time, so we can use it to represent the evolution
of the worldsheet of the defects in different regions by mapping to this coordinate system.

In order to match the space times across the bubble light cone we take $-\pi \le \eta \le 0$, $0 \le \chi \le \pi$ for the false vacuum and  $\chi-\pi/2 \le \eta \le \arcsin\left( \cos \chi \tanh \left|\log\gamma\right|\right)$, $0 \le \chi \le \pi/2$ for the true vacuum. The conformal coordinates relate to the flat slicing coordinates that cover the false vacuum through
\ba
&&e^{-H_F t} =-\frac{\sin \eta}{\cos \eta + \cos \chi},\nonumber \\
&&r=\frac{\sin \chi}{\cos \eta + \cos \chi},
\ea
and relate to the open slicing coordinates that cover the true vacuum through 
\ba
&&\sinh H_T \tau = -\frac{1}{\sinh\left(\widetilde{\eta} + \log\gamma \right)}~,~~\tanh \widetilde{\eta} = \frac{\sin \eta}{\cos \chi},\nonumber \\
&&\tanh \xi = \frac{\sin \chi}{\cos \eta}.
\ea


\section{Jacobian Matrix}
\label{app:JM}

In this Appendix, we calculate the Jacobian of the transformation matrix from $(r_n, c)$ to $(\xi_m, \Delta\phi)$. To simplify the calculation, we define
\ba
M=e^{\xi_m},~P=\cos\left(\frac{\Delta\phi}{2}\right) .
\ea
Then we have
\ba
c' = \frac{(1+M) (1+P+M (P-1))}{1+M^2+P-M^2 P}
\ea
and
\ba
r_n'=\frac{2 M}{(1+M) (1+P+M (P-1))},
\ea
And according to the mapping (\ref{apcp}), we have
\ba
&&r_n =\frac{2 M \gamma }{1-M^2 (1-P)+P+2 M P \gamma }, \nonumber\\
&&c=\frac{1+M^2 (-1+P)+P+2 M P \gamma }{\sqrt{\left(1+M^2 (-1+P)+P\right)^2-4 M^2 \left(-1+P^2\right) \gamma ^2}}.
\ea
Thus,
\ba
\left|\frac{\partial\left(r_n,c\right)}{\partial\left(M,P\right)}\right|=\frac{4 M \left(1-M^2 (-1+P)+P\right) \gamma ^2}{\left(\left(1+M^2 (-1+P)+P\right)^2-4 M^2 \left(-1+P^2\right) \gamma ^2\right)^{3/2}} ,
\ea
and finally we obtain
\ba
\left|\frac{\partial\left(r_n,c\right)}{\partial\left(\xi_m,\Delta\phi\right)}\right| = \frac{4 M \left(1-M^2 (-1+P)+P\right) \gamma ^2}{\left(\left(1+M^2 (-1+P)+P\right)^2-4 M^2 \left(-1+P^2\right) \gamma ^2\right)^{3/2}} \frac{dM}{d\xi_m}\frac{dP}{d\Delta\phi}.
\ea


\section{Total number of loops within a comoving radius $r$}
\label{app:ncs}

In this appendix, we outline the calculation of the integral~(\ref{Nr}). We first define new variables, $A= c^2r_n^2$ and $C=c^2r^2$.  Then
\ba
dN&&= 4 \pi \lambda  \sin \alpha_m c^2 r_n^2 dr_n dc,\\
&&=\frac{\lambda\pi}{C} \sqrt{A-\frac{1}{4}\left(A+1-C\right)^2} dA d C,
\ea
and the integration domain is bounded by
\ba
\sqrt{A}>\frac{\sqrt{C}}{r}-1,~~\sqrt{A}<1+\sqrt{C},~~c>0,~~\sqrt{A}>1-\sqrt{C}.
\ea
One can verify that $0<\sqrt{C}<\frac{2r}{1-r}$ for $1-\sqrt{C} < \sqrt{A} <1+\sqrt{C}$, and $\frac{2r}{1+r}<\sqrt{C}<\frac{2r}{1-r}$ for $1+\sqrt{C}<\sqrt{A}<\sqrt{C}/r-1$. We also separate the $A$-integration over $1+\sqrt{C}<\sqrt{A}<\sqrt{C}/r-1$ into two parts, with a separation point at $A=1+C$.  Then the number of loops $N(r)$ is expressed as a sum of three integrals,
\ba
N(r) &&= \lambda\pi \int_0^{\frac{4r^2}{(1-r)^2}} dC \int_{(1-\sqrt{C})^2}^{(1+\sqrt{C})^2} dA \sqrt{A-\frac{1}{4}\left(A+1-C\right)^2}/C \\
&&+~\lambda\pi\int_{\frac{4r_b^2}{(1+r)^2}}^{\frac{4r^2}{(1-r)^2}} dC \int_{1+C}^{(1+\sqrt{C})^2} dA\sqrt{A-\frac{1}{4}\left(A+1-C\right)^2}/C \\
&&+~\lambda\pi\int_{\frac{4r^2}{(1+r)^2}}^{\frac{4r^2}{(1-r)^2}} dC \int_{(\sqrt{C}/r-1)^2}^{1+C} dA\sqrt{A-\frac{1}{4}\left(A+1-C\right)^2}/C.
\ea
Performing the integrations, we obtain
\be
N(r)=\frac{\lambda\pi^2r^2(r^2+16r+4)}{(1-r^2)^2}.
\ee


\section{Collision marks on the sky}
\label{app:skymap}
In this Appendix we shall describe the method we used to generate the distribution of collision marks on sky.

\subsection{Domain walls}

For a domain wall centered on the positive part of the $x$-axis, the intersection of the worldsheets of the wall and the bubble wall can be found from the equations
\ba
&&\left\{
\begin{aligned}
&r^2-2r_n r \sin \theta \cos \phi + r_n^2 =e^{-2t_n} + e^{-2t}\\
&r=1-e^{-t}
\end{aligned}
\right. .
\ea
The past light cone of the centered observer intersects the bubble worldsheet at $r=r_l(\tau_O, \gamma)$,
where $\tau_O$ is the time elapsed from bubble nucleation to the observation point. Using the coordinates defined in Appendix \ref{app:coordinates}, we find $r_l =\frac12  \left(1+\tanh\frac{\widetilde\eta_O}{2}\right)$ with $\sinh H_T \tau_O=-\frac{1}{\sinh (\widetilde{\eta}_O+\log\gamma)}$. In the limit of $\tau_O \gg H_T^{-1}$, we have 
\be
r_l\approx \frac12 \left(1-\tanh\frac{\log\gamma}{2}\right),
\ee 
which reduces to $r_l \approx \gamma^{-1}$ for $\gamma\ll 1$. The observable part of the intersection is a disc-like region bounded by 
\ba
r=\frac{r_n^2-e^{-2t_n}-1}{2(r_n\sin\theta\cos\phi-1)},
\ea
with $r \le r_l$, that is
\ba\label{dwshape}
2(1-r_n \sin \theta \cos \phi)r_l  =e^{-2t_n}-r_n^2+1.
\ea

If the center of the domain wall is not on the $x$-axis, the outline of the resulting collision mark on the sky can be found from Eq.~(\ref{dwshape}) by applying a suitable rotation.

For domain walls with nucleation centers in the blue region of the parameter space, the angular extent of this region on the sky is $\Delta\phi = 2\phi_m$, where 
\ba
\phi_m=\arccos\left[\frac{r_n^2e^{2t_n}+e^{2t_n}(2r_l-1)-1}{2r_n r_l e^{2t_n}}\right] =\arccos\left[\frac{(r_l-1)\cos\chi_n+r_l\cos\eta_n}{r_l\sin\chi_n}\right].
\ea
For nucleation centers in the green region, the angular extent is $\Delta\phi= 2\pi-2\phi_m$. ``Blue" domain  walls always have $\Delta\phi<\pi$.  This means that the area affected by the collision is less than half of the sky.  For green walls, $\Delta\phi$ can either be greater or smaller than $\pi$, but we find numerically that collisions with $\Delta\phi > \pi$ are much more likely.

\begin{figure}
\centering
 \includegraphics[width=0.5\linewidth]{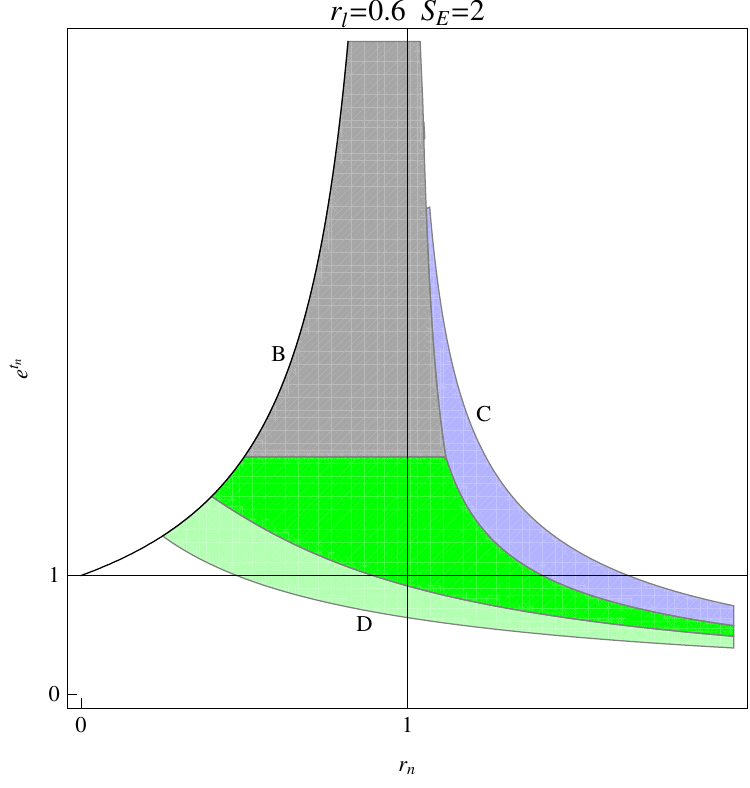}%
\caption{The parameter space of domain walls with $r_l=0.6$ and $S_E=2$. Both green and blue regions continue to infinity in the direction of $r_n$.}
\label{fig:dwps}
\end{figure}

An example of the parameter space $(c=e^{t_n},r_n)$ for $r_l=0.6$, $S_E=2$ is shown in Fig.~\ref{fig:dwps}.  As before, nucleation centers in the blue area correspond to walls hitting the bubble from the right, and nucleation centers in green areas (both dark and light) correspond to walls hitting the bubble from the left.  The dark and light green regions correspond to collisions with $\Delta\phi>\pi$ and $\Delta\phi <\pi$, respectively.  The curve $B$ is the bubble boundary.

For a collisions to be observable, we require that $t_c<t_l$, which gives
\ba
c<\frac{1}{\sqrt{(r_n-1)(r_n-2r_l+1)}}
\ea
for blue walls and
\ba
c > \frac{1}{\sqrt{(r_n+1)(r_n+2r_l-1)}}
\ea
for green walls.  The resulting cutoff curves are labeled $C$ and $D$ in the figure.

We also require $t_f < t_c $ for blue walls and $t_f<0$ for green walls.  The corresponding constraints are, respectively, 
\ba
r_n>1-\frac{S_E-\sqrt{1+S_E^2}}{c}
\ea 
and $c<S_E$.  They exclude the gray region of the parameter space in the figure. 

The blue and green areas of the parameter space extend to infinite values of $r_n$, but the volume of the spacetime region that they specify is finite.  We randomly pick parameter sets $(t_n, r_n)$ from the allowed region according to the distribution (\ref{crn})
\be
dN = 4\pi \lambda r_n^2 dr_n c^2 d c.
\label{Nrnc}
\ee
For each parameter set we choose a random direction and plot the resulting collision mark on the sky with the Mollweide projection.

\subsection{Cosmic Strings}

The intersection between the worldsheets of a string loop centered on the $x$-axis and the bubble wall can be found from the equations
\ba\label{csintersection}
\left\{
\begin{aligned}
&2(1-r_n \sin \theta \cos \phi)r  =e^{-2t_n}-r_n^2+1 \\
&\left(\sin\theta\cos\phi\tan\alpha-  \cos \theta \right)r=r_n\tan\alpha
\end{aligned}
\right.~.
\ea
The projection of the intersection line on the celestial sphere can be easily expressed in a parametric form $\theta(r)$, $\phi(r)$; we do not reproduce the explicit expressions here. An observer at $r=0$ can see only the part of this line at $r\leq r_l$.  If the center of the loop is not on the $x$-axis, the resulting collision mark on the sky can be found by applying a suitable rotation.

As before, we impose the conditions $t_f < t_c < t_l$ for all loops and $t_f<0$ for loops with nucleation centers in the green region. We then use the distribution (\ref{Nrncalpha}) to randomly choose the nucleation parameters $(t_n ,r_n,\alpha)$ in this restricted parameter space.  Finally, we randomly pick a direction to the loop nucleation center and plot the collision mark with the Mollweide projection.


\section{Distribution of collision events for a non-central observer}
\label{app:persistence}

We shall now study collision events within the past light cone of a non-central observer. It will be convenient to introduce flat-slicing coordinates inside the bubble,
\ba
ds^2 = dt'^2 - H_T^{-2} e^{2H_Tt'} \left(dx'^2+dy'^2+dz'^2 \right),
\ea
in terms of which the past light cone of an observer at $(t_0', x_0',0,0)$ is
\ba
|\vec{x}'-\vec{x}'_0|=e^{-H_Tt'}-e^{-H_T t_0'}=\sqrt{(x'-x_0')^2+y'^2+z'^2},
\ea
and the bubble wall is
\ba
|\vec{x}'| =1- e^{-H_Tt'}=\sqrt{x'^2+y'^2+z'^2}.
\ea
The intersection of the past light cone and the bubble wall can be found by combining the above equations,
\ba
&&e^{-H_Tt_e'}=\frac12\frac{x_0'^2+1-e^{-2H_Tt_0'}-2x_0'\cos \psi}{1-e^{-H_Tt_0'}-x_0'\cos\psi} \\
&&r_e'=\frac12\frac{(1-e^{-H_Tt_0'})^2-x_0'^2}{1-e^{-H_Tt_0'}-x_0'\cos \psi},
\ea
where we have introduced $\cos \psi = x'/r'$ and $r'=|\vec{x}'|$. In the limit of $t_0' \to \infty$, the spatial projection of the intersection surface is an ellipsoid
\ba
r_e'=\frac{1-x_0'^2}{2(1-x_0' \cos \psi)},
\ea
which has one focus at the origin and the other at the observer's location $\vec{x}'_0$.  Note that here $r_e'$ is a comoving coordinate.

In order to find the number of string loops that intersect with this surface, we should first rewrite the ellipsoid in terms of the flat-slicing coordinates outside the bubble. The junction between the interior and exterior spacetimes requires that the physical radii of two-spheres match at the location of the wall, namely
\ba
H_F^{-1}e^{H_Ft}r=H_T^{-1}e^{H_Tt'}r'.
\ea
For points on the bubble wall, this gives
\ba
r_e=\frac{r_e'}{(1-\gamma)r_e'+\gamma}.
\ea

According to Ref.~\cite{Basu}, nucleating string loops are uniformly distributed on a constant $t$ surface.  The number of loops of physical radius $R$ whose centers are located in a 3-volume element $dV=2\pi r_{\text{phy}}^2  \sin \psi dr_{\text{phy}} d\psi$ is
\ba\label{dNcopy}
dN = \lambda \frac{dR}{R^4} dV .
\ea
This distribution is cut off at $R \approx H_F^{-1}$. 

In order to intersect with the ellipsoid, a loops of radius $R\ll H_T^{-1}$ should have its center in the range
\ba
H_T^{-1}\frac{r_e'}{1-r_e'}+H_F^{-1} < r_{\text{phy}} < H_T^{-1}\frac{r_e'}{1-r_e'}+R,
\ea
where we used $e^{-H_Tt_e'}=1-r_e'$. Then the distribution of intersections with respect to $\psi$ is
\ba\label{disce1}
\frac{d{\cal N}}{d\psi} &&\simeq 2\pi \lambda  \sin \psi \int_{H_F^{-1}}^{\sim H_T^{-1}} \frac{dR}{R^4} \int_{\frac{r_e'}{1-r_e'}H_T^{-1}+H_F^{-1}}^{ \frac{r_e'}{1-r_e'}H_T^{-1}+R}  r_{\text{phy}} ^2 dr_{\text{phy}} \\
&& \simeq \frac{\pi}{3} \lambda \gamma^{-2} \sin \psi \frac{r_e'^2}{(1-r_e')^2}\\
&&=\frac{\pi}{3} \lambda\gamma^{-2}  \frac{\left(x_0'^2-1\right)^2 \sin \psi}{\left(1+x_0'^2-2 x_0' \cos  \psi \right)^2},
\ea
where we have evaluated the integrals in the limit of $\gamma\ll 1$.

Introducing the zenith angle $\psi_O$ with respect to the non-central observer, the above distribution becomes
\ba\label{disce2}
\frac{d{\cal N}}{d\psi_O} \simeq 2\pi \lambda \gamma^{-2} \sin \psi_O.
\ea
We thus see that there is no anisotropy in the number distribution of collision events in the limit of $\gamma \ll 1$. The total number of collision events within the past light cone can be found by integrating Eq.~(\ref{disce2}) over $\psi_O$,
\ba\label{Nce}
{\cal N}=\frac{2\pi}{3} \lambda \gamma^{-2},
\ea
which is the same as (\ref{calN}).

\end{document}